\documentclass[review]{elsarticle}

\usepackage{lineno,hyperref}
\modulolinenumbers[5]

\usepackage{bm} \usepackage{amsmath} \usepackage{amsfonts}
\usepackage{xcolor} \usepackage{graphicx} \usepackage{inputenc}
\usepackage[margin=1.2in]{geometry}
\usepackage[section]{placeins}
\usepackage{float}
\usepackage{array}
\usepackage{datetime}
\usepackage{enumerate}
\usepackage{nomencl} \makenomenclature

\journal{Journal of Power Sources}

\begin{document}

\textbf{Highlights:}

\begin{itemize}
  \item A hybrid PF model is proposed to study fracture behavior of electrode during lithiation.
  \item The PF model is applied to a Si NW electrode.
  \item The concentration-dependent elastic properties is taken into account.
  \item Hybrid model shows less tendency to crack growth than isotropic model.
\end{itemize}

\begin{frontmatter}

\title{A Hybrid Phase Field Model for Fracture Induced by Lithium Diffusion in Electrode Particles of Li-ion Batteries}

%% include affiliations in footnotes:
\author{Masoud Ahmadi} \ead{masoudahmadi@msc.guilan.ac.ir}

\address{Faculty of Mechanical Engineering, University of Guilan, P.O. Box 3756, Rasht, Iran}

\begin{abstract}
Lithium-ion batteries (LIBs) of high energy density and light-weight design, have found wide applications in electronic devices and systems. Degradation mechanisms that caused by lithiation is a main challenging problem for LIBs with high capacity electrodes like silicon (Si), which eventually can reduce the lifetime of batteries. In this paper, a hybrid phase field model (PFM) is proposed to study the fracture behavior of LIB electrodes. The model considers the coupling effects between lithium (Li) -ion diffusion process, stress evolution and crack propagation. Also, the dependency of Elastic properties on the concentration magnitude of Li-ion is considered. A numerical implementation based on a MATLAB finite element (FE) code is elaborated. Then, the proposed hybrid PF approach is applied to a Nanowire (NW) Si electrode particle. It is shown that the hybrid model shows less tendency to crack growth than the isotropic model.
\end{abstract}

\begin{keyword}
Phase field model\sep Hybrid formulations\sep Lithium-ion batteries\sep Crack propagation\sep Finite element method\sep Electrode particles.
\end{keyword}

\end{frontmatter}

%% Nomenclature
\nomenclature{$F$}{Total free energy}
\nomenclature{$c$}{Li concentration field}
\nomenclature{$\phi$}{Fracture order field}
\nomenclature{$\bm u$}{Displacement field}
\nomenclature{$f_c$}{Local free energy density of $c$}
\nomenclature{$f_\phi$}{Local free energy density of $\phi$}
\nomenclature{$f_u$}{Local free energy density of $\bm u$}
\nomenclature{$k_c$}{Gradient energy coefficient of $c$}
\nomenclature{$k_\phi$}{Gradient energy coefficient of $\phi$}
\nomenclature{$\nabla$}{Nabla operator}
\nomenclature{$\cdot$}{Dot product}
\nomenclature{$:$}{Double dot product}
\nomenclature{$\square^T$}{Transpose of a matrix}
\nomenclature{$\xi_u$}{Elastic energy density function}
\nomenclature{$\xi_u^+$}{Tension part of the elastic energy density function}
\nomenclature{$\xi_u^-$}{Compression part of the elastic energy density function}
\nomenclature{$\mu_0$}{Standard chemical potential}
\nomenclature{$N_A$}{Avogadro's constant}
\nomenclature{$k_B$}{Boltzmann constant}
\nomenclature{$T$}{Absolute temperature}
\nomenclature{$c_{\textrm{max}}$}{Maximum Li concentration}
\nomenclature{$G_{\textrm{cr}}$}{Critical energy release rate}
\nomenclature{$l_0$}{Regularization constant}
\nomenclature{$g(\phi)$}{Energetic degradation function}
\nomenclature{$g'(\phi)$}{Derivative of energetic degradation function with respect to $\phi$}
\nomenclature{$\bm{\sigma}$}{Cauchy stress tensor/vector}
\nomenclature{$\bm{\varepsilon}$}{Infinitesimal Strain tensor/vector}
\nomenclature{$\eta$}{Threshold stiffness parameter for a fully broken region}
\nomenclature{$\bm{\varepsilon}_c$}{Chemical strain tensor}
\nomenclature{$\bm{\varepsilon}_e$}{Elastic strain tensor}
\nomenclature{$\Omega$}{Partial molar volume}
\nomenclature{\textbf{I}}{Second-order unit tensor}
\nomenclature{$\mathbb{D}$}{Fourth-order stiffness tensor}
\nomenclature{$\sigma_p$}{Hydrostatic stress}
\nomenclature{$\bar{f}$}{Integral term of $F$}
\nomenclature{$M$}{Molecular mobility}
\nomenclature{$\chi$}{Relaxation constant of the fracture order}
\nomenclature{$\textbf{J}$}{Ion flux vector}
\nomenclature{$\bar{\sigma}_1$}{Largest principal value of the effective stresses}
\nomenclature{$\langle \square \rangle$}{Macaulay brackets}
\nomenclature{$\delta$}{Variation operator}
\nomenclature{$\partial$}{Partial differential operator}
\nomenclature{$\textrm{d}$}{Total differential operator}
\nomenclature{$\eta_c$}{Test function for $c$}
\nomenclature{$\eta_\phi$}{Test function for $\phi$}
\nomenclature{$\eta_{\bm u}$}{Test function for $\bm u$}
\nomenclature{$\dot{\square}$}{First material time derivative}
\nomenclature{$\ddot{\square}$}{Second material time derivative}
\nomenclature{$m$}{Number of nodes of each element}
\nomenclature{$N^I$}{Nodal values of shape functions}
\nomenclature{$\bm{B_u}^I$}{Usual strain matrix}
\nomenclature{$\bm{B_c}^I$}{Cartesian derivative matrix}
\nomenclature{$\bm{B_\phi}^I$}{Cartesian derivative matrix}
\nomenclature{$\bm{D_1}$}{2D elasticity matrix}
\nomenclature{$\bm{D_2}$}{A constant vector}
\nomenclature{$\bm{D_3}$}{A constant vector}
\nomenclature{$D_4$}{A constant scalar}
\nomenclature{$E$}{Young's modulus}
\nomenclature{$\nu$}{Poisson's ratio}
\nomenclature{$\bm{Q}$}{Field variables vector}
\nomenclature{$\beta$}{A constant scalar}
\nomenclature{$\gamma$}{A constant scalar}
\nomenclature{$\Delta$}{Increment}
\nomenclature{$n$}{Time step index}
\nomenclature{$t$}{Time}
\nomenclature{$i$}{Newton iteration index}
\nomenclature{$\bm R$}{Nodal residual vector}
\nomenclature{$\bm K$}{Tangential stiffness matrix}
\nomenclature{$\bm D$}{Damping matrix}
\nomenclature{$V$}{Volume}
\nomenclature{$\ln$}{Natural logarithm}
\nomenclature{$I$}{Node index}
\nomenclature{$R$}{Radius of NW electrode}
\nomenclature{$a$}{Initial crack length}
\nomenclature{$h$}{Thickness of NW electrode}
\nomenclature{$c_0$}{Initial Li concentration}

\printnomenclature

\section{Introduction}\label{sec.Intro}

Due to high energy storage density, LIBs are widely used in different technologies such as portable electronic devices and electric vehicles \cite{smith2010electrochemical}. The working principle of LIB cells lies essentially in the electrochemical potential driven redox reaction in the electrode active materials \cite{zhao2019review}. In LIBs, Li-ions transfer from the anode and diffuse through the electrolyte towards the cathode during charge and when the battery is discharged, the respective electrodes change roles. Abundant efforts have been made to develop next-generation battery systems with novel electrode materials, however, most candidates with promising electrochemical potential have chemo-mechanical stability problems \cite{mukhopadhyay2014deformation}. Due to Graphite low volume changes (about $10 \%$) under intercalation, Graphite is a common anode among the commercial batteries \cite{kumar2018carbon}. Silicon is a promising candidate material with a theoretical capacity of $4200$ mA~h/g, which is exceedingly higher than the theoretical capacity of Graphite with $372$ mA~h/g. Nevertheless, Si shows a volume changes of about $300 \%$, which causes high stress in electrodes that eventually lead to fracture, fragmentation, or pulverization \cite{obrovac2004structural, ebner2013visualization, zhao2019review}. Cracking of electrodes under diffusion is one of the main reasons for the short life span of LIBs with high capacity electrodes \cite{zuo2015phase}.

For a better understanding of the fracture behavior of LIBs during lithiation, numbers of modeling and computational simulations have been expanded, which can offer insights into the structural reliability of electrodes in LIBs. Ryu et al. \cite{ryu2011size} used a fracture mechanics approach to study fracture behavior of Si NWs during lithiation/delithiation process. They considered large deformation associated with lithiation and effects of pressure gradients on the diffusion of Li in their study. Grantab and Shenoy \cite{grantab2012pressure} proposed a method that accounts for the effects of pressure-gradients within the material on the flux for studying lithiation-induced crack propagation in Si NWs. Their model consists of FE simulations in which the pressure-gradients are computed numerically to capture the effect of the crack-tip on the localized diffusion. Chen et al. \cite{chen2016analytical} established an analytical model to study the stress evolution and crack propagations in spherical particle electrodes during phase transformation. Hu et al. \cite{hu2017surface} developed an analytical model to study the diffusion induced stress evolution and discussed the crack growth by using stress intensity factor coupled with surface effects. They showed that smaller Si Spherical particles exhibit higher structural integrity. Wang et al. used Peridynamic theory to study the lithiation induced stress and fracture in Si square thin film electrodes \cite{wang2018predicting, wang2018peridynamic}, and Si spherical and cylindrical nanoparticles \cite{wang2018three}. Gwak et al. \cite{gwak2018cohesive} investigated the stress evolution and dynamic cracking behavior of Si NWs using a chemo-mechanical model based on the large deformation theory and the bilinear cohesive zone model.

In sharp-interface modeling of phase-separating behavior, a discontinuity is imposed at the interface, and a subsequent strain mismatch will give rise to stress concentration at the interface region \cite{hu2010averting}. Phase field modeling as one of the most prominent approaches for the simulation of microstructure evolution is introduced as an alternative to sharp-interface modeling. PFM is based on the thermodynamic description of non-equilibrium states in materials including interfaces and is formulated as a state variable in space and time, the evolution of which controls the pathway towards equilibrium \cite{steinbach2009phase}. Phase field fracture models were developed by regularizing the sharp crack topology by a diffuse damage band through the introduction of phase field parameter that discriminate the intact and broken material \cite{bourdin2000numerical, bourdin2008variational, aranson2000continuum, karma2001phase, hakim2009laws, spatschek2011phase, lu2019crack}. Phase field fracture models have been employed in order to investigate the fracture behavior of the electrodes in LIBs during lithiation/delithiation process. Zuo and Zhao \cite{zuo2015phase} developed a PFM coupling Li diffusion and stress evolution with crack propagation to investigate the behavior of Si thin film electrodes. Miehe et al. \cite{miehe2016phase} studied chemo-mechanical induced fracture in LIBs by PF fracture modeling. Zuo et al. \cite{zuo2016phase} proposed a PFM coupling Li diffusion, finite deformation, stress evolution with crack propagation to study spherical Si electrodes in LIBs. Guan et al. \cite{guan2018phase} studied the stress evolution and crack propagation in the SEI layer formed on the LiMn$_2$O$_4$ electrode during the Li-ion diffusion by using a PFM.

In the isotropic PF fracture models, cracking may arise in regions under compression, thus leading to unphysical crack growth patterns. All the PF fracture models that have been discussed yet are isotropic models. In order to prevent cracking under compressive load states, a tension/compression split of the strain energy is proposed in the so-called anisotropic models \cite{amor2009regularized, miehe2010thermodynamically}. In some PF simulations of lithiation induced stress and fracture in electrode particles in the literature such as \cite{zhao2016phase, zhang2016variational, klinsmann2016modeling2, klinsmann2016modeling3, klinsmann2016modeling, xu2016phase, nguyen2018phase}, the split of the strain energy into tensile and compressive parts is considered. By using PFM of fracture coupled with anisotropic Cahn-Hilliard-type diffusion, Zhao et al. \cite{zhao2016phase} investigated the electrochemical reaction in LIBs. A chemo-mechanical coupled computational framework was formulated by Zhang et al. \cite{zhang2016variational} to model diffusion induced large plastic deformation and PF fracture of Si electrodes. Klinsmann et al. developed a coupled model of Li diffusion, mechanical stress and crack growth, that uses a PF method for the fracture to investigate crack growth in LiMn$_2$O$_4$ spherical and cylindrical particles during Li extraction \cite{klinsmann2016modeling2} and insertion\cite{klinsmann2016modeling3} in the first half cycle and in the 2nd half cycle \cite{klinsmann2016modeling}. Xu et al. \cite{xu2016phase} investigated the fracture behavior of cylinder electrode particles by using a finite strain PF fracture model, in which phase segregation and electrochemical reaction on both particle surfaces and fracture surfaces have been taken into account. Nguyen et al. \cite{nguyen2018phase} developed a new formulation based on the PF method for modeling stress corrosion cracking induced by anodic dissolution. They coupled classical phase transition model for material dissolution with the mechanical problem and applied their model to an aluminum alloy in a saline medium.

From the computational viewpoint, anisotropic models are significantly more expensive than the isotropic ones, since the tension/compression split of the strain energy leads to nonlinear balance of momentum equations \cite{ambati2015review}. To overcome this issue, the so-called hybrid PFM has been proposed in recent years \cite{wu2017unified, wu2018geometrically, ambati2015review, wu2018length}. The hybrid formulation formally comprises features from both the isotropic and anisotropic models \cite{ambati2015review}. In this paper, the author proposes a FE-based hybrid PF model to study the fracture behavior of electrode particle in LIBs during lithiation. The proposed model with coupling effects among Li diffusion, stress evolution and crack propagation, enables a significant reduction of computational cost in comparison with the available anisotropic models.

\section{Phase field model}\label{sec.PFM}

In this section, a PF model with hybrid formulation based on the total free energy of the system is developed, which is characterized by elastic deformation, Li concentration and crack propagation.

\subsection{Free energy functional}\label{sec.PFM.FEF}
The total free energy of the system is formulated as a functional of the parameters, $\bm u$, $c$ and $\phi$ \cite{zuo2016phase}
\begin{equation}\label{eq.freeenergy}
F=\int \left( f_u + f_c + f_\phi + \frac{1}{2} \ k_c \ (\nabla c)^2 + \frac{1}{2} \ k_\phi \ (\nabla \phi)^2  \right) \textrm{d}V \, ,
\end{equation}
where $f_u$, $f_c$ and $f_\phi$ are the local free energy densities of the displacement field $\bm u$, Li concentration field $c$ and fracture order field $\phi$, and $\nabla c$ and $\nabla \phi$ are the gradients of concentration and fracture respectively. Also, $k_c$ and $k_\phi$ represent the gradient energy coefficients.

Considering the coupling between the elastic field and the fracture field, $f_u$ is described as follows
\begin{equation}\label{eq.fumodeling}
f_u = \frac{1}{2} \ g(\phi) \ \bm{\sigma} : \bm{\varepsilon} = g(\phi) \ \xi_u \, ,
\end{equation}
where $\xi_u = \frac{1}{2} \ \bm{\sigma} : \bm{\varepsilon}$ indicates the elastic energy density function, $\bm{\sigma}$ denotes the Cauchy stress tensor and $\bm{\varepsilon}$ is the infinitesimal strain tensor which is given by
\begin{equation}\label{eq.straindisplacement}
\bm{\varepsilon}= \frac{1}{2} \left( \nabla \bm u + (\nabla \bm u)^T \right)\, .
\end{equation}
The energetic degradation function $g(\phi)$ couples the elastic field and fracture order parameter by considering the stiffness loss between an intact ($\phi = 1$) and a fully broken region ($\phi = 0$). The $g(\phi)$ has to be monotonically decreasing and also satisfies $g(0)=0$, $g(1)=1$, and $\partial g(0)/ \partial \phi=0$. Herein, the fourth quartic polynomial function $g(\phi)=4 \ \phi^3 - 3 \ \phi^4 + \eta$ is chosen from the several available degradation functions in the literature \cite{ambati2015review,kuhn2015degradation}. The parameter $\eta$ is a residual stiffness for a fully broken region and its value may not be too small because of stability issues \cite{kuhn2010continuum}. The influence of the concentration field on the elastic field is modeled in analogy to thermal effects \cite{prussin1961generation,li1978physical}. Therefore, the total infinitesimal strain in the material is decomposed into chemical and elastic parts as
\begin{equation}\label{eq.straindecompose}
\bm{\varepsilon}= \bm{\varepsilon}_e + \bm{\varepsilon}_c\, ,
\end{equation}
where $\bm{\varepsilon}_e$ is the elastic strain and $\bm{\varepsilon}_c$ is the chemical strain which is assumed as a hydrostatic dilatation as $\bm{\varepsilon}_c=(\Omega~c) /3 \ \textbf{I}$, where $\Omega$ is the partial molar volume and \textbf{I} is the second-order unit tensor. The constitutive equation for the mechanical stresses is given by
\begin{equation}\label{eq.stressconcentration}
\bm{\sigma} = \mathbb{D} : \bm{\varepsilon}_e = \mathbb{D} : \left( \bm{\varepsilon} - c \ \frac{\Omega}{3} \ \textbf{I} \right) \, . \\
\end{equation}
Therein, $\mathbb{D}$ is the fourth-order stiffness tensor.

In general, $f_c$ is modeled by a regular solution model \cite{singh2008intercalation} or an ideal solution model \cite{yang2005interaction}. Here, for $f_c$ it holds \cite{yang2005interaction}
\begin{equation}\label{eq.fcmodeling}
\frac{\partial f_c}{\partial c} = \mu_0 + N_A \ k_B \ T \ c_{\textrm{max}} \ \ln c \, ,
\end{equation}
where the constant parameter $\mu_0$, is the standard chemical potential, $N_A$ is Avogadro's constant, $k_B$ is the Boltzmann constant, $c_{\textrm{max}}$ is the maximum Li concentration, and $T$ is the absolute temperature.

From the \cite{miehe2010thermodynamically,miehe2010phase}, $f_\phi$ is modeled as follows
\begin{equation}\label{eq.fphimodeling}
f_\phi = \frac{G_{\textrm{cr}}}{2 \ l_0} \ (1-\phi)^2 \, ,
\end{equation}
where, $G_{\textrm{cr}}$ is the critical energy release rate and $l_0$ controls the width of the transition area between the unbroken and the broken region. Also, the $k_\phi$ parameter in Eq. \eqref{eq.freeenergy}, is assumed to be $k_\phi = G_{\textrm{cr}} \ l_0$.

In the present multi-field modeling, the following coupling effects are considered:
\begin{itemize}
  \item Li diffusion affects elastic field by the chemical strain $\bm{\varepsilon}_c$.
  \item Fracture affects elastic field by the energetic degradation function $g(\phi)$.
\end{itemize}
It should be noted that no direct coupling between Li concentration and fracture is considered in this model. Nevertheless, the coupling effects among Li diffusion, elastic field and crack propagation are captured via bridging interactions, i.e. Li concentration can cause elastic deformation which can lead to crack propagation.

\subsection{Strong form of equations}\label{sec.PFM.SF}
For each variation of \eqref{eq.freeenergy} the Cahn-Hilliard equation and the Ginzburg-Landau equation can be derived through a variational method \cite{chen2002phase}. Due to the characteristic time of the elastic field being far less than the concentration field and fracture field, the evolution equation of displacement is assumed to be quasi-static. The governing equations of the system are derived by solving the Ginzburg-Landau equations for the evolution of locally non-conserved parameters $\bm u$ and $\phi$, and the Cahn-Hilliard equation for the evolution of locally conserved parameter $c$, \cite{zuo2015phase}
\begin{align}\label{eq.GinzburgLandau1}
\frac{\partial \bm u}{\partial t} &= - \frac{\delta F}{\delta \bm u} = \nabla \frac{\partial \bar{f}}{\partial (\nabla \bm u)} - \frac{\partial \bar{f}}{\partial \bm u} = 0 \, , \\
\label{eq.GinzburgLandau2}
\frac{\partial \phi}{\partial t} &= - \chi \ \frac{\delta F}{\delta \phi} = \chi \left( \nabla \frac{\partial \bar{f}}{\partial (\nabla \phi)} - \frac{\partial \bar{f}}{\partial \phi} \right) \, , \\
\label{eq.Cahn.Hilliard}
\frac{\partial c}{\partial t} &= \nabla M \ \nabla \left( \frac{\delta F}{\delta c} \right) \, ,
\end{align}
where $\bar{f}$ is the integral term of the total free energy of the system in \eqref{eq.freeenergy}, and $M$ and $\chi$ are molecular mobility and relaxation constant of the fracture order filed respectively.

Equation \eqref{eq.GinzburgLandau1} is the balance of linear momentum which can be written explicitly as
\begin{equation}\label{eq.equilibrium}
\nabla \cdot (g(\phi) \ \bm{\sigma}) =0 \, .
\end{equation}

According to \eqref{eq.freeenergy} and \eqref{eq.fcmodeling}, \eqref{eq.Cahn.Hilliard} finally yields the mass conservation equation which can be written explicitly as follows
\begin{equation}\label{eq.massconservation}
\frac{\partial c}{\partial t} + \nabla \cdot \textbf{J} = 0\, ,
\end{equation}
where $\textbf{J}$ is the ion flux vector given by
\begin{equation}\label{eq.fluxvector}
\textbf{J} = -M k_B \ T \left[ \nabla c - \left( \frac{\Omega}{N_A \ K_B \ T} \right) c \ \nabla ( g(\phi) \ \sigma_p ) \right] \, ,
\end{equation}
with hydrostatic stress $\sigma_p$, in turn defined by
\begin{equation}\label{eq.sigmah}
\sigma_p = \frac{1}{3} \ \bm{\sigma}:\textbf{I} \, .
\end{equation}
So, the evolution of the Li concentration can be written in a more convenient form by inserting \eqref{eq.fluxvector} into \eqref{eq.massconservation} as \cite{zuo2015phase}
\begin{equation}\label{eq.lievolution}
\frac{\partial c}{\partial t} - M k_B \ T \ \nabla^2 c + \frac{M c \ k_c}{N_A \ c_{\textrm{max}}} \nabla^4 c + \frac{M k_c}{N_A \ c_{\textrm{max}}} \nabla \nabla^2 c \cdot \nabla c + \frac{M c \ \Omega}{N_A} \nabla^2 ( g(\phi) \ \sigma_p ) + \frac{M \Omega}{N_A} \nabla c \cdot \nabla ( g(\phi) \ \sigma_p )=0 \, .
\end{equation}
Following \cite{xie2015phase,zuo2015phase}, for convenience in the numerical simulations, the fourth order terms in \eqref{eq.lievolution} are neglected in numerical study in Sec.~\ref{sec.FEM}.

By inserting Eq. \eqref{eq.freeenergy} and \eqref{eq.fphimodeling} to Eq. \eqref{eq.GinzburgLandau2}, the crack propagation equation becomes
\begin{equation}\label{eq.fipropagation}
\frac{1}{\chi} \frac{\partial \phi}{\partial t} + G_{\textrm{cr}} \ l_0 \ \nabla^2 \phi + \frac{G_{\textrm{cr}}}{l_0} (1-\phi) - g'(\phi) \ \xi_u =0 \, .
\end{equation}
The $g'(\phi)$ is the derivative of energetic degradation function with respect to $\phi$.

Equations \eqref{eq.equilibrium}, \eqref{eq.lievolution} and \eqref{eq.fipropagation} are the strong form of governing equations of the electrode for isotropic PF model.

\subsection{Hybrid formulations}\label{sec.PFM.HYB}
The isotropic models allow for cracking in regions under compression and interpenetration of the crack faces, hence it yields physically unrealistic crack evolution patterns. To address this issue, the anisotropic models suggest a tension/compression split of elastic energy density function, i.e.,
\begin{equation}\label{eq.anisotropic}
f_u = \ g(\phi) \ \xi_u^+ + \xi_u^- \, ,
\end{equation}
such that \eqref{eq.fipropagation} turns into
\begin{equation}\label{eq.fipropagation2}
\frac{1}{\chi} \frac{\partial \phi}{\partial t} + G_{\textrm{cr}} \ l_0 \ \nabla^2 \phi + \frac{G_{\textrm{cr}}}{l_0} (1-\phi) - g'(\phi) \ \xi_u^+ =0 \, ,
\end{equation}
where the energetic degradation function $g(\phi)$ is only applied to the tension part of the elastic energy density function.

Although the anisotropic formulation prevent crack growth in compressive regions, it leads to nonlinear balance of momentum equations, hence makes the numerical treatment more expensive \cite{ambati2015review}. The general idea of hybrid (isotropic-anisotropic) formulation is to retain a linear momentum balance equation and also prevent crack growth in regions under compressive load states. To this end, the not separated form of $f_u$ in \eqref{eq.fumodeling} is kept from isotropic model, while \eqref{eq.fipropagation} is replaced by \eqref{eq.fipropagation2} from anisotropic model which contains tension part of the elastic energy density. Therefore, Equations \eqref{eq.equilibrium}, \eqref{eq.lievolution} and \eqref{eq.fipropagation2} are the complete set of governing equations for hybrid PF model. The only seeming disadvantage of the hybrid formulation is that it is variationally inconsistent \cite{ambati2015review}, however, it does not violate the second law of thermodynamics \cite{wu2018robust}.

The tension part of the elastic energy density function, $\xi_u^+$ is given by \cite{wu2018robust}
\begin{equation}\label{eq.xiup}
\xi_u^+ = \frac{1}{2E} \ \langle \bar{\sigma}_1 \rangle^2 \, ,
\end{equation}
where $E$ is Young's modulus, $\bar{\sigma}_1$ denotes the largest principal value of the effective stresses which is described in \ref{app.sig1b}, and Macaulay brackets $\langle \square \rangle$ are defined as $\langle x \rangle=\textrm{max} \{x,0\}$.

\subsection{Weak form of equations}\label{sec.PFM.WF}
Utilizing the tests functions $\eta_c$ for $c$, $\eta_\phi$ for $\phi$, $\bm{\eta_u}$ for $\bm u$, and integrate over the domain, results in the weak forms of \eqref{eq.equilibrium}, \eqref{eq.lievolution} and \eqref{eq.fipropagation2} as
\begin{align}\label{eq.weak.u}
&\int\limits_V \nabla \bm{\eta_u} : (g(\phi) \ \bm{\sigma}) \ \textrm{d}V = 0 \, , \\
\label{eq.weak.c}
&\int\limits_V \dot{c} \ \eta_c \ \textrm{d}V
+ \int\limits_V (M k_B \ T) \ \nabla c \cdot \nabla \eta_c \ \textrm{d}V - \int\limits_V \left( \frac{M c \ \Omega}{N_A} \right) \ \nabla ( g(\phi) \ \sigma_p ) \cdot \nabla \eta_c \ \textrm{d}V = 0 \, , \\
\label{eq.weak.phi}
&\int\limits_V \frac{1}{\chi} \ \dot{\phi} \ \eta_\phi \ \textrm{d}V
+ \int\limits_V G_{\textrm{cr}} \ l_0 \ \nabla \phi \cdot \nabla \eta_\phi \ \textrm{d}V - \int\limits_V \frac{G_{\textrm{cr}}}{l_0} (1-\phi) \ \eta_\phi \ \textrm{d}V + \int\limits_V g'(\phi) \ \xi_u^+ \ \eta_\phi \ \textrm{d}V = 0 \, ,
\end{align}
where $\dot{c}$ and $\dot{\phi}$ are the material time derivative of the Li concentration field and fracture order parameter respectively. It should be noted that, the external nodal forces applied on the electrode boundaries are considered in the global residual vector by means of the FE code.

\section{Finite element implementation}\label{sec.FEM}

In the implementation of numerical method, matrix-vector notation is more convenient than tensor notation. Hence, we switch to the matrix-vector notation for the rest of this paper. In matrix-vector notation, a second-order symmetric tensor is expressed using a vector, while a fourth-order symmetric tensor is expressed using a matrix.

Since the numerical example considered in the present work can be accurately and effectively be treated by a two-dimensional plane strain model. So, the $\bm{\sigma, \, \varepsilon} \, \textrm{and} \,  \bm u$ from \eqref{eq.stressconcentration} and \eqref{eq.straindisplacement} are defined in vector form as
\begin{equation}\label{eq.vectorform}
\bm{\sigma} = \left[ \sigma_{xx},\ \sigma_{yy},\ \sigma_{xy} \right]^T, \quad \bm{\varepsilon} = \left[ \varepsilon_{xx},\ \varepsilon_{yy},\ 2\varepsilon_{xy} \right]^T, \quad \bm u = \left[ u_x,\ u_y \right]^T \, .
\end{equation}

\subsection{Spatial discretisation}\label{sec.FEM.SD}
By taking nodal variables as $\bm{u}$, $c$ and $\phi$, and utilizing the element shape functions, \eqref{eq.weak.u}, \eqref{eq.weak.c} and \eqref{eq.weak.phi} are discretized by isoparametric elements
\begin{align}\label{eq.dis.u}
\bm u = \sum_{I}^{m} N^I \ \bm{u}^I, \qquad \bm{\varepsilon} = \sum_{I}^{m} \bm{B_u}^I \ \bm{u}^I \, , \\
\label{eq.dis.c}
c = \sum_{I}^{m} N^I \ c^I, \qquad \nabla c = \sum_{I}^{m} \bm{B}_c^I \ c^I \, , \\
\label{eq.dis.phi}
\phi = \sum_{I}^{m} N^I \ \phi^I, \qquad \nabla \phi = \sum_{I}^{m} \bm{B}_\phi^I \ \phi^I \, ,
\end{align}
where $m$ is the number of nodes of each element and $N^I$ is the nodal values of shape functions. The $\bm{B_u}^I$ is the usual strain matrix, and $\bm{B}_c^I$ and $\bm{B}_\phi^I$ are the Cartesian derivative matrices, which are expressed as
\begin{equation}\label{eq.B.cu}
\bm{B_u}^I = \left[
          \begin{array}{cc}
            \frac{\partial N^I}{\partial x} & 0 \\
            0 & \frac{\partial N^I}{\partial y} \\
            \frac{\partial N^I}{\partial y} & \frac{\partial N^I}{\partial x} \\
          \end{array}
        \right]
 , \qquad \bm{B}_c^I = \bm{B}_\phi^I = \left[
                    \begin{array}{cc}
                      \frac{\partial N^I}{\partial x} \\
                      \frac{\partial N^I}{\partial y}
                    \end{array}
                  \right] \, .
\end{equation}
Without losing the generality of formulation, test functions can be expressed as the variations of different fields, namely
\begin{align}\label{eq.dis.eu}
\bm{\eta_u} = \sum_{I}^{m} N^I \ \delta \bm{u}^I, \qquad \nabla \bm{\eta_u} = \sum_{I}^{m} \bm{B_u}^I \ \delta \bm{u}^I \, , \\
\label{eq.dis.ec}
\eta_c = \sum_{I}^{m} N^I \ \delta c^I, \qquad \nabla \eta_c = \sum_{I}^{m} \bm{B}_c^I \ \delta c^I \, , \\
\label{eq.dis.ephi}
\eta_\phi = \sum_{I}^{m} N^I \ \delta \phi^I, \qquad \nabla \eta_\phi = \sum_{I}^{m} \bm{B}_\phi^I \ \delta \phi^I \, .
\end{align}
So, the variational formulations of \eqref{eq.weak.u}, \eqref{eq.weak.c} and \eqref{eq.weak.phi} are achieved as
\begin{align}\label{eq.var.u}
&\int\limits_V \left( \delta \bm{u}^I \right)^T \ \left( \bm{B_u}^I \right)^T g(\phi) \ \bm{\sigma} \ \textrm{d}V = 0 \, , \\
\label{eq.var.c}
&\int\limits_V N^I \ \dot{c} \ \delta c^I \ \textrm{d}V
+ \int\limits_V (M k_B \ T) \ (\bm{B}_c^I)^T \ \nabla c \ \delta c^I \ \textrm{d}V - \int\limits_V \left( \frac{M \Omega}{N_A} \right) \ \left( \bm{B}_c^I \right)^T \ \nabla (g(\phi) \ \sigma_p) \ c \ \delta c^I \ \textrm{d}V = 0 \, , \\
\label{eq.var.phi}
&\int\limits_V \frac{1}{\chi} \ \dot{\phi} \ N^I \ \delta \phi^I \ \textrm{d}V
+ \int\limits_V G_{\textrm{cr}} \ l_0 \ (\bm{B}_\phi^I)^T \ \nabla \phi \ \delta \phi^I \ \textrm{d}V - \int\limits_V \frac{G_{\textrm{cr}}}{l_0} \ N^I \ (1-\phi) \ \delta \phi^I \ \textrm{d}V \nonumber \\ &+ \int\limits_V N^I \ g'(\phi) \ \xi_u^+ \ \delta \phi^I \ \textrm{d}V = 0 \, .
\end{align}

Stress vector and hydrostatic stress from \eqref{eq.stressconcentration} and \eqref{eq.sigmah} can be decomposed as
\begin{align}\label{eq.stressconcentration2}
\bm{\sigma} = \bm{D}_1 \ \bm{\varepsilon} + \bm{D}_2 \ c = \bm{D}_1 \ \bm{B_u}^I \ \bm{u}^I + \bm{D}_2 \ N^I \ c^I \, , \\
\label{eq.sigmah2}
\sigma_p = \frac{1}{3} \ \left[ \left( \bm{D}_3 \right)^T \ \bm{\varepsilon} + \textrm{D}_4 \ c \right] = \frac{1}{3} \ \left[ \left( \bm{D}_3 \right)^T \ \bm{B_u}^I \ \bm{u}^I + \textrm{D}_4 \ N^I \ c^I \right] \, .
\end{align}
where $\bm{D}_1$ is the $3 \times 3$ elasticity matrix in plane-stress or plane-strain cases. Furthermore, $\bm{D}_2$, $\bm{D}_3$ and $\textrm{D}_4$ for plane-stress assumption are expressed as
\begin{equation}\label{eq.Ds1}
\bm{D}_2 = - \frac{E \Omega}{3(1- \nu)} \ [1,1,0]^T, \ \textrm{D}_3 = \frac{E}{1- \nu}  \ [1,1,0]^T, \ \textrm{D}_4 = - \frac{2 \ E \Omega}{3(1- \nu)} \, ,
\end{equation}
and for plane-strain assumption are expressed as
\begin{equation}\label{eq.Ds2}
\bm{D}_2 = - \frac{E \Omega}{3(1-2 \nu)} \ [1,1,0]^T, \ \textrm{D}_3 = \frac{E}{1-2 \nu}  \ [1,1,0]^T, \ \textrm{D}_4 = - \frac{E \Omega}{1-2 \nu} \, ,
\end{equation}
in which $\nu$ is Poisson's ratio. So, $\nabla (g(\phi) \ \sigma_p)$ can be calculated as
\begin{equation}\label{eq.grsigmah}
\nabla (g(\phi) \ \sigma_p) = g(\phi) \ \nabla \sigma_p + \nabla g(\phi) \ \sigma_p = g(\phi) \ \left( \frac{1}{3} \ \left[ \left( \bm{D}_3 \right)^T \ \nabla \bm{\varepsilon} + \textrm{D}_4 \ \nabla c \right] \right) + \left(12 \ \phi^2 (1-\phi) \ \nabla \phi \right) \ \sigma_p \, .
\end{equation}
For convenience in differentiation, $\left( \bm{D}_3 \right)^T \ \nabla \bm{\varepsilon} = \bm{D}_3^* \ \nabla \bm{\varepsilon}^* $, which $\bm{D}_3^*$ for plane-stress and plane-strain respectively is
\begin{equation}\label{eq.dstar}
\bm{D}_3^* = \frac{E}{1- \nu} \left[ \begin{array}{cccc}
1 & 1 & 0 & 0 \\ 0 & 0 & 1 & 1 \\ \end{array} \right], \ \bm{D}_3^* = \frac{E}{1-2\nu} \left[ \begin{array}{cccc}
1 & 1 & 0 & 0 \\ 0 & 0 & 1 & 1 \\ \end{array} \right] \, ,
\end{equation}
and
\begin{equation}\label{eq.ebstars}
\nabla \bm{\varepsilon}^* = \left[
                              \begin{array}{c}
                                u_{x,xx} \\
                                u_{y,yx} \\
                                u_{x,xy} \\
                                u_{y,yy} \\
                              \end{array}
                            \right] = \sum_{I}^{n} {\bm{B_u}^I}^* \ u_i^I,
\qquad {\bm{B_u}^I}^* = \left[
                      \begin{array}{cc}
                            \frac{\partial^2 N^I}{\partial x^2}& 0 \\
                             0 & \frac{\partial^2N^I}{\partial y \partial x} \\
                           \frac{\partial^2N^I}{\partial x \partial y} & 0 \\
                           0 & \frac{\partial^2 N^I}{\partial y^2} \\
                         \end{array}
                    \right] \, .
\end{equation}

\subsection{Temporal discretization}\label{sec.FEM.TD}
The field variables vector is defined as $\bm Q = \{ \bm u, c, \phi \}^T$, hence based on the quasi-static assumption for the displacement, the first and second material time derivative vectors are defined as $\dot{\bm Q}= \left\{\bm 0, \dot{c}, \dot{\phi} \right\}^T$ and $\ddot{\bm Q}= \left\{\bm 0, \ddot{c}, \ddot{\phi} \right\}^T$. An implicit time integration method, a Newmark method with time step size $\Delta t^{n+1} = t^{n+1} - t^n$ is used for temporal discretization, where the following approximations for $\dot{\bm Q}$  and $\ddot{\bm Q}$ at time $t^{n+1}$ are considered \cite{newmark1959method,wriggers2008nonlinear},
\begin{align}\label{eq.qdot}
\dot{\bm Q}^{n+1} = \frac{\gamma}{\beta \ \Delta t^{n+1}} \left( {\bm Q}^{n+1} - {\bm Q}^{n} \right) + \left( 1 - \frac{\gamma}{\beta} \right) \dot{\bm Q}^{n} + \left( 1 - \frac{\gamma}{2 \ \beta} \right) \Delta t^{n+1} \ \ddot{\bm Q}^{n} \, , \\
\label{qddot}
\ddot{\bm Q}^{n+1} = \frac{1}{\beta \ \left( \Delta t^{n+1} \right)^2 } \left( {\bm Q}^{n+1} - {\bm Q}^{n} \right) - \frac{1}{\beta \ \Delta t^{n+1}} \ \dot{\bm Q}^{n} - \frac{1-2 \ \beta}{2 \ \beta} \ \ddot{\bm Q}^{n} \, ,
\end{align}
where $\gamma$ and $\beta$ are constants parameters. Considering both accuracy and stability, $\gamma$ and $\beta$ are set to be $\gamma = \beta = 0.5$. For more details about possible choices of $\gamma$ and $\beta$, see \cite{wriggers2008nonlinear,zienkiewicz2000finite}.

%The field variables vector is defined as $\bm Q = \{ \bm u, c, \phi \}^T$, hence based on the quasi-static assumption for the displacement, the material time derivative vector is defined as $\dot{\bm Q}= \left\{\bm 0, \dot{c}, \dot{\phi} \right\}^T$. An implicit time integration method, a backward Euler method with time step size $\Delta t^{n+1} = t^{n+1} - t^n$ is used for temporal discretization. Therefore, the material time derivative of concentration and crack fields are expressed as
%\begin{equation}\label{eq.cdotphidot}
%\dot{c}=\frac{c^{n+1}-c^n}{\Delta t^{n+1}}, \qquad \dot{\phi}=\frac{\phi^{n+1}-\phi^n}{\Delta t^{n+1}} \, .
%\end{equation}

\subsection{Solving system of nonlinear equations of electrode}\label{sec.FEM.NE}
The nonlinear set of equations of system, \eqref{eq.var.u}, \eqref{eq.var.c} and \eqref{eq.var.phi}, can be written in the following form
\begin{equation}\label{eq.algeqsy}
\bm{R}=\bm{C}(\bm{\dot{Q}}^{n+1})+\bm{P}(\bm{Q}^{n+1})=0 \, .
\end{equation}
The incremental equation system, Newton method is applied to determine the unknown vector $\bm{Q}^{n+1}$. This leads to the iterative scheme with iteration index $i$,
\begin{align}\label{eq.itersch}
\left[ \frac{\gamma}{\beta \ \Delta t^{n+1}} \ \bm{D} + \bm{K}(\bm{Q}^{n+1}_i) \right] \Delta \bm{Q}^{n+1}_{i+1} &= -\bm R(\bm{Q}^{n+1}_i) \, , \\
\bm{Q}^{n+1}_{i+1} &= \bm{Q}^{n+1}_i + \Delta \bm{Q}^{n+1}_{i+1} \, ,
\end{align}
which will be calculated at each time step. The nodal residual vector, $\bm R^{I} = \left\{ \bm{R}^{I}_{\bm{u}}, R^{I}_c, R^{I}_\phi \right\}^T$, can be expressed as
\begin{subequations}\label{eq.res}
\begin{align}
\bm{R}^{I}_{\bm{u}} &= \int\limits_V \left( \bm{B_u}^I \right)^T g(\phi) \ \bm{\sigma} \ \textrm{d}V \, , \\
R^{I}_c &= \int\limits_V (M k_B \ T) \ (\bm{B}_c^I)^T \ \nabla c \ \textrm{d}V - \int\limits_V \left( \frac{M \Omega}{N_A} \right) \ \left( \bm{B}_c^I \right)^T \ \nabla (g(\phi) \ \sigma_p) \ c \ \textrm{d}V \, , \\
R^{I}_\phi &= \int\limits_V G_{\textrm{cr}} \ l_0 \ (\bm{B}_\phi^I)^T \ \nabla \phi \ \textrm{d}V - \int\limits_V N^I \ \left( \frac{G_{\textrm{cr}}}{l_0} \ (1-\phi) - g'(\phi) \ \xi_u^+ \right) \ \textrm{d}V \, .
\end{align}
\end{subequations}
The tangential stiffness matrix $\bm{K}$ and damping matrix $\bm{D}$ can be derived as
\begin{equation}\label{eq.tangdamp}
\bm{K}^{IJ} = \frac{\partial \bm{P}^{I}}{\partial \bm{Q}^{J}} \, , \quad \bm{D}^{IJ} = \frac{\partial \bm{C}^{I}}{\partial \bm{Q}^{J}} \, ,
\end{equation}
\begin{equation}\label{eq.tangdamp2}
\bm{K}^{IJ} = \left[
                  \begin{array}{ccc}
                    \bm{K}^{IJ}_{\bm{uu}} & \bm{K}^{IJ}_{\bm{u}c} & \bm{K}^{IJ}_{\bm{u}\phi} \\
                    \bm{K}^{IJ}_{c \bm{u}} & K^{IJ}_{cc} & K^{IJ}_{c\phi} \\
                    \bm{K}^{IJ}_{\phi \bm{u}} & K^{IJ}_{\phi c} & K^{IJ}_{\phi \phi} \\
                  \end{array}
                \right]
, \qquad \bm{D}^{IJ} = \left[
                  \begin{array}{ccc}
                    0 & 0 & 0 \\
                    0 & D^{IJ}_{cc} & 0 \\
                    0 & 0 & D^{IJ}_{\phi \phi} \\
                  \end{array}
                \right] \, .
\end{equation}
The nodal contributions to the $\bm{K}^{IJ}$ matrix result from
\begin{subequations}\label{eq.stif}
\begin{align}
\bm{K}^{IJ}_{\bm{uu}} = \frac{\partial \bm{P}^{I}_{\bm{u}}}{\partial \bm{u}^J} &= \int\limits_V g(\phi) \ \left( \bm{B_u}^I \right)^T \bm{D}_1 \ \bm{B_u}^J \ \textrm{d}V \, , \\
\bm{K}^{IJ}_{\bm{u}c} = \frac{\partial \bm{P}^{I}_{\bm{u}}}{\partial c^J} &= \int\limits_V g(\phi) \ \left( \bm{B_u}^I \right)^T \bm{D}_2 \ N^J \ \textrm{d}V \, , \\
\bm{K}^{IJ}_{\bm{u} \phi} = \frac{\partial \bm{P}^{I}_{\bm{u}}}{\partial \phi^J} &= \int\limits_V g'(\phi) \ \left( \bm{B_u}^I \right)^T \bm{\sigma} \ N^J \ \textrm{d}V \, , \\
\bm{K}^{IJ}_{c\bm{u}} = \frac{\partial P^{I}_c}{\partial \bm{u}^J} &= \int\limits_V - \left( \frac{M \Omega}{N_A} \right) \ (\bm{B}_c^I)^T \left[ g(\phi) \left( \frac{1}{3} \bm{D}_3^* \ {\bm{B_u}^J}^* \right) + (12 \ \phi^2 (1-\phi) \ \nabla \phi ) \left( \frac{1}{3} (\bm{D}_3)^T {\bm{B_u}^J} \right) \right] \ c \ \textrm{d}V \, , \\
K^{IJ}_{cc} = \frac{\partial P^{I}_c}{\partial c^J} &= \int\limits_V (M k_B \ T) \ (\bm{B}_c^I)^T \ \bm{B}_c^J \ \textrm{d}V - \int\limits_V \left( \frac{M \Omega}{N_A} \right) \ (\bm{B}_c^I)^T \ \nabla (g(\phi) \ \sigma_p) \ N^J \textrm{d}V \nonumber \\ &- \int\limits_V \left( \frac{M \Omega}{N_A} \right) \ (\bm{B}_c^I)^T \ \left[ g(\phi) \left( \frac{1}{3} \ \textrm{D}_4 \ \bm{B}_c^J \right) + (12 \ \phi^2 (1-\phi) \ \nabla \phi ) \left( \frac{1}{3} \ \textrm{D}_4 \ N^J \right) \right] \ c \ \textrm{d}V \, , \\
K^{IJ}_{c \phi} = \frac{\partial P^{I}_c}{\partial \phi^J} &= \int\limits_V \left( \frac{M \Omega}{N_A} \right) \ (\bm{B}_c^I)^T \ \left[ g'(\phi) \ \nabla \sigma_p + \sigma_p \ (12 \ \phi^2 (1-\phi) \ \bm{B}_\phi^J + 6 \phi \ (4+6\phi) \ \nabla \phi \ N^J ) \right] \ c \ \textrm{d}V \, , \\
\bm{K}^{IJ}_{\phi \bm{u}} = \frac{\partial P^{I}_\phi}{\partial \bm{u}^J} &= \int\limits_V N^I \ \left( \frac{\partial \left(g'(\phi) \ \xi_u^+ \right)}{\partial \bm{u}} \right) \ \textrm{d}V \, , \\
K^{IJ}_{\phi c} = \frac{\partial P^{I}_\phi}{\partial c^J} &= \int\limits_V N^I \ \left( \frac{\partial \left(g'(\phi) \ \xi_u^+ \right)}{\partial c} \right) \ \textrm{d}V \, , \\
K^{IJ}_{\phi \phi} = \frac{\partial P^{I}_\phi}{\partial \phi^J} &= \int\limits_V G_{\textrm{cr}} \ l_0 \ (\bm{B}_\phi^I)^T \ \bm{B}_\phi^J \ \textrm{d}V + \int\limits_V N^I \ \left( \frac{G_{\textrm{cr}}}{l_0} + g''(\phi) \ \xi_u^+ \right) \ N^J \ \textrm{d}V \, .
\end{align}
\end{subequations}
For calculating $\partial \left(g'(\phi) \ \xi_u^+ \right) / \partial \bm{u}$ and $\partial \left(g'(\phi) \ \xi_u^+ \right) / \partial c$ in \ref{eq.stif}g and \ref{eq.stif}h, the readers are referred to \ref{app.tpeedfd}. The nodal contributions to the $\bm{D}^{IJ}$ matrix result from
\begin{subequations}\label{eq.damp}
\begin{align}
D^{IJ}_{cc} = \frac{\partial C^{I}_c}{\partial c^J} &= \int\limits_V N^I \ N^J \ \textrm{d}V \, , \\
D^{IJ}_{\phi \phi} = \frac{\partial C^{I}_\phi}{\partial \phi^J} &= \int\limits_V \frac{1}{\chi} \ N^I \ N^J \ \textrm{d}V \, .
\end{align}
\end{subequations}

After forming element stiffness, damping and residuals, they are assembled into the global stiffness matrix, damping matrix and the right-hand side vector. The resulting set of nonlinear equation have been implemented in Matlab. The implementation of the isotropic model is presented in \ref{app.isomi}.

\subsection{Irreversibility constraints}\label{sec.PFM.IRR}
Crack propagation is an irreversible process. However, the system of equations does not guarantee the irreversible evolution of non-conserved PF fracture field, where crack healing is possible. To prevent crack from healing, irreversibility constraints are considered. Miehe et al. \cite{miehe2010thermodynamically, miehe2010phase} proposed a local history field of the maximum tension part of the elastic energy density function, which automatically deals with the irreversibility conditions. Nevertheless, this irreversibility condition cannot be used in the present PFM since the boundedness cannot be guaranteed for non-quadratic energetic degradation functions. Herein, a damage like formulation is utilized, where $\Delta \phi = - \langle - \Delta \phi \rangle$ ensures the irreversibility constraint.

\section{Numerical example}\label{sec.Numerical}
\begin{figure}[H]
\begin{center}
\begin{tabular}{c}
  \includegraphics[scale=0.7]{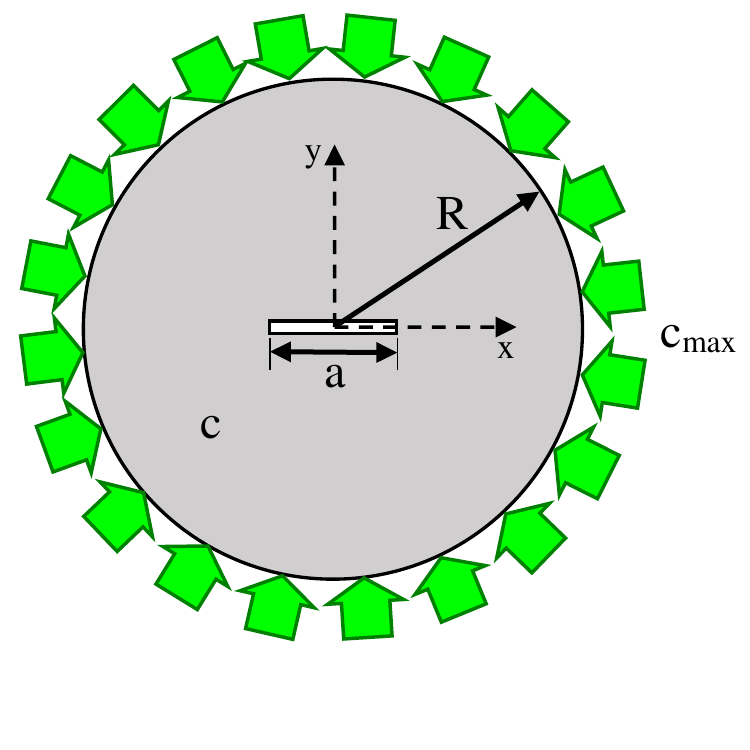} \\
\end{tabular}
\caption{Geometry of and loading conditions on the NW cross section plate.}\label{fig.bc}
\end{center}
\end{figure}

To study the crack propagation under diffusion, the hybrid and also isotropic PF simulation of a Si NW electrode is explored. Two-dimensional circular plate is considered for modeling the cross section of NW electrode as shown in Fig. \ref{fig.bc}. The radius of the circle plate is equal to $R = 60~$nm and its thickness is $h= 3~\mu$m \cite{toriyama2002single, li2016nanostructured}. Also, there is an initial crack with a length of $a$, which is located in the center of the specimen plate. The plane strain assumption is conducted and the edges of the models are free from any displacement constraints. As shown a constant Li concentration, $c_{\textrm{max}} = 88.67$~Kmol/m$^3$ \cite{ryu2011size} is applied on boundaries of the electrode. Also, the initial value of concentration in the inner part of electrode is $c_0=c~(t=0)=1.0$~Kmol/m$^3$ \cite{xie2015phase}. As the time progresses, Li-ions start to diffuse from outer boundaries into the inner region of the electrode particle.

The material properties of Si electrode \cite{xie2015phase, pharr2013measurements, berla2015mechanical, shenoy2010elastic} are listed in Tab.~\ref{tab.pro}. Unless otherwise stated, the fracture properties are set to be $G_{\textrm{cr}}=7~$N/m and $l_0=10~$nm, and the length of pre-existing crack is considered to be $a=60~$nm.
\begin{table}[htbp]
	\caption{Material properties of Si electrode \cite{xie2015phase,pharr2013measurements,berla2015mechanical,shenoy2010elastic}.}
	\label{tab.pro}
	\centering
	\begin{tabular}{l c c c}
		\hline
		Parameter & Symbol & Value & Unit \\
		\hline
		Young's modulus of Si & $E_{\textrm{Si}}$ & 80 & GPa \\
		Poisson's ratio of Si & $\nu_{\textrm{Si}}$ & 0.22 & - \\
        Young's modulus of Li--Si & $E_{\textrm{Li--Si}}$ & 41 & GPa \\
        Poisson's ratio of Li--Si & $\nu_{\textrm{Li--Si}}$ & 0.24 & - \\
		Partial molar volume & $\Omega$ & $8.5 \times 10^{-6}$ & m$^3$/mol \\
		Molecular mobility & $M$ & $500$ & m$^2$/Js \\
		Boltzmann constant & $k_B$ & $1.38 \times 10^{-23}$ & J/K \\
		Absolute temperature & $T$ & 298.15 & K \\
		Avogadro's constant & $N_A$ & $6.02 \times 10^{23}$ & 1/mol \\
        Relaxation constant & $\chi$ & $1.25 \times 10^{-10}$ & m$^3$/Js \\
        Critical energy release rate & $G_{\textrm{cr}}$ & $2.4-14.9$ & N/m \\
        Length parameter for crack & $l_0$ & $5-30$ & nm \\
		\hline
	\end{tabular}
\end{table}

\subsection{Concentration-dependent elastic properties}\label{sec.NUM.CDEP}
Elastic properties of Li--Si alloys strongly depend on the Li concentration \cite{zhao2016phase, ratchford2011young}. Herein, based on a linear rule of mixtures, $E = E_{\textrm{Li--Si}} + (1-c/c_{\textrm{max}}) (E_{\textrm{Si}}-E_{\textrm{Li--Si}})$, the dependency of Young's modulus on concentration of lithated Si is considered. Also, with a similar relationship, $\nu = \nu_{\textrm{Li--Si}} + (1-c/c_{\textrm{max}}) (\nu_{\textrm{Si}}-\nu_{\textrm{Li--Si}})$, the dependency of Poisson's ratio on Li concentration is considered.

\subsection{Mesh and time step convergence}\label{sec.NUM.Mesh}
Choosing a suitable mesh size is critical to obtain sound results at low cost and appropriate time span. For meshing the model, 4-node isoparametric quadrilateral elements with the refined mesh around the vicinity of the crack are employed. Elaborate mesh convergence analyses are conducted to ensure the full convergence of nonlinear FE solution. Besides, time step convergence analyses show that for $\Delta t \geq 0.006~$s the solution process shows divergence and for $\Delta t \leq 0.003~$s the results are virtually converged in time. The $\Delta t = 0.0025~$s temporal discretization is selected for the next simulations in the paper.

\subsection{Effect of initial crack length}\label{sec.NUM.hybiso}
\begin{figure}[H]
	\begin{center}
		\begin{tabular}{c c}
			\hspace{3em} {\large Isotropic} \hspace{4em} \ {\large Hybrid} & {\large Isotropic} \hspace{4em} \ {\large Hybrid} \\
           \includegraphics[scale=0.22]{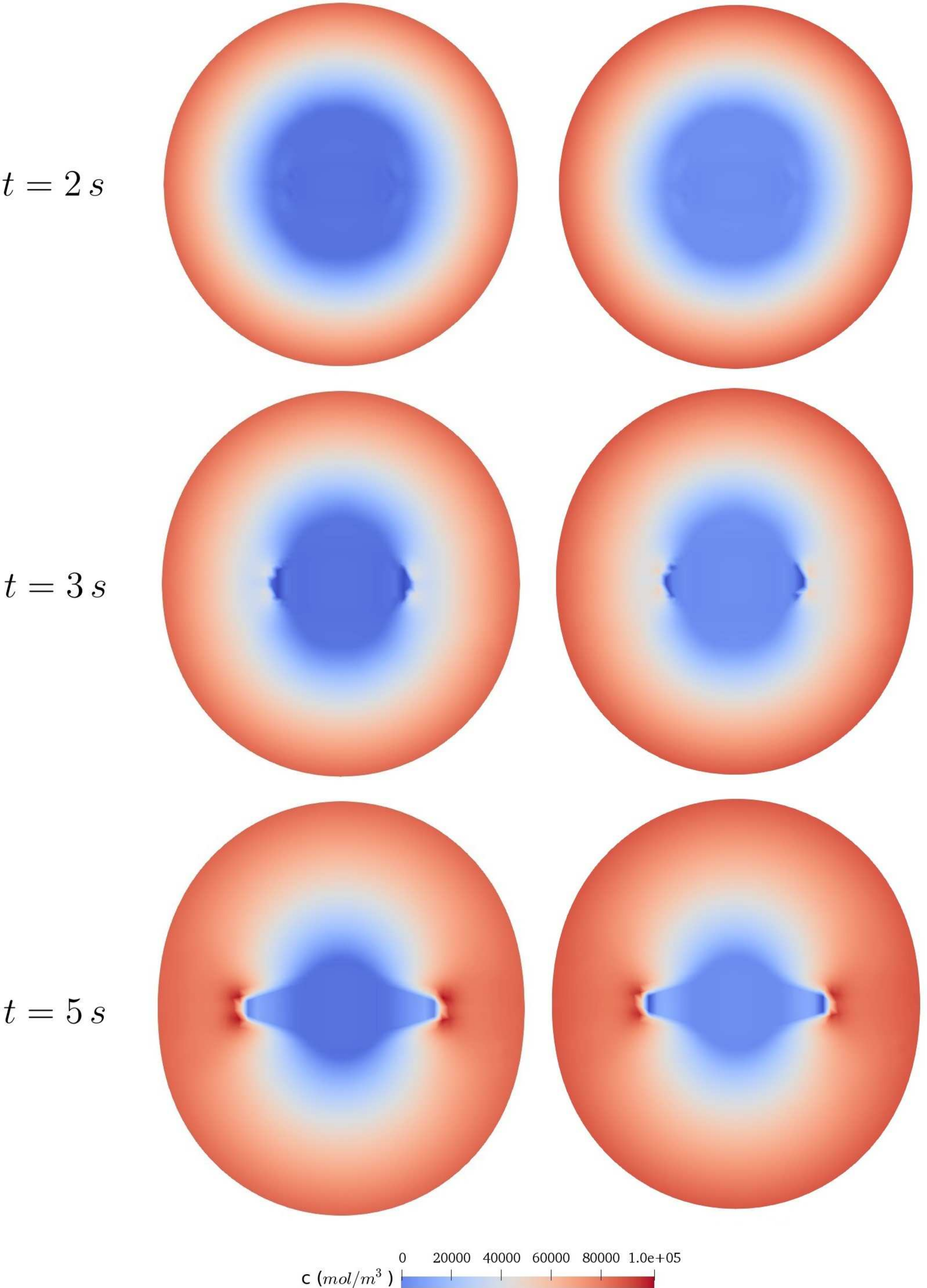} & \includegraphics[scale=0.22]{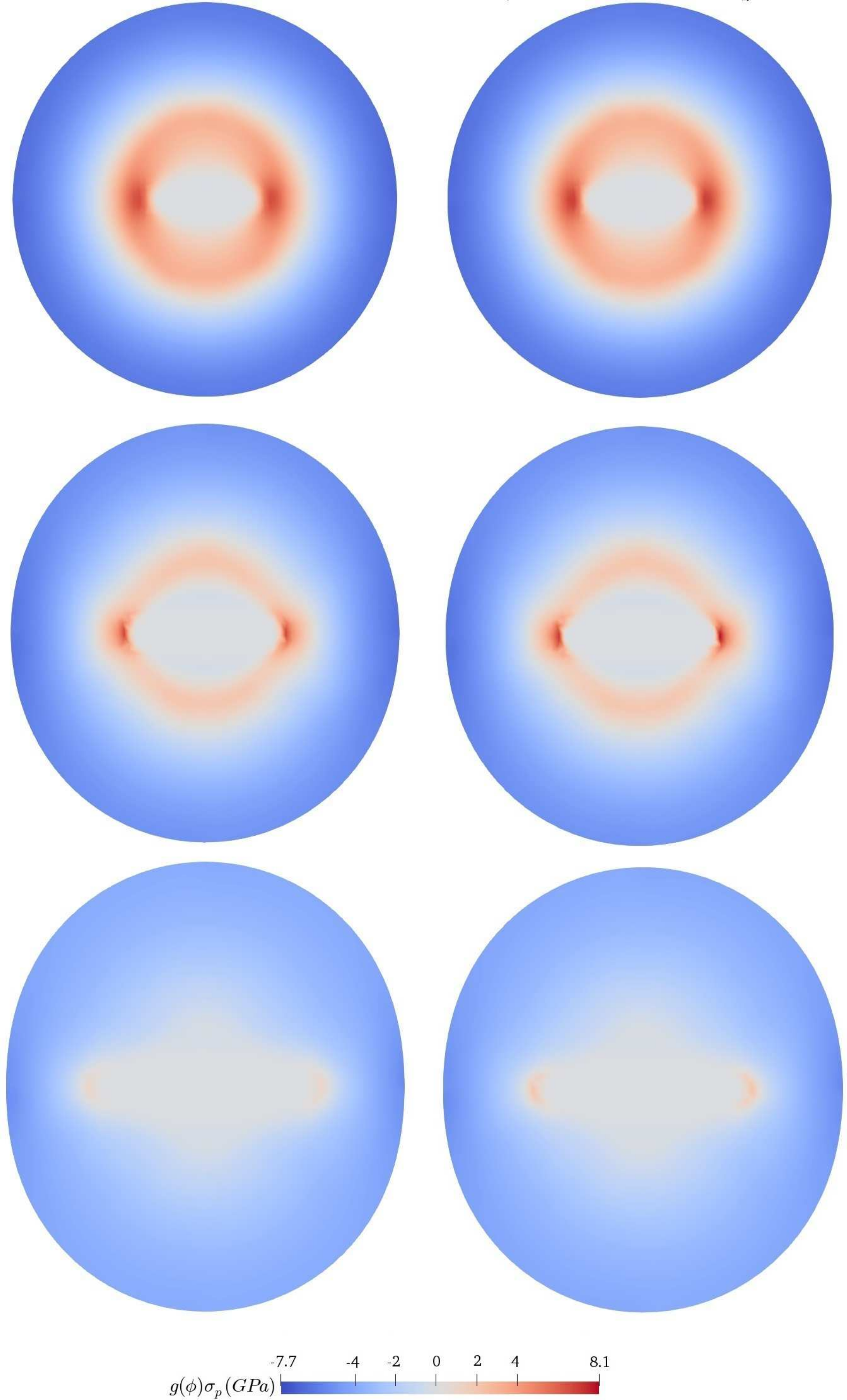} \\
		\end{tabular}
		\caption{The evolutions of $c$ and $g(\phi)\sigma_p$ distributions for isotropic and hybrid models.}\label{fig.cont-c}
	\end{center}
\end{figure}

The evolutions of $c$ and $g(\phi)\sigma_p$ distributions in the electrode during the charging process for the isotropic and hybrid model are depicted in Fig. \ref{fig.cont-c}. It can be seen that at the beginning of process, $c$ increases radially from the boundaries into the center of the plate and the interface gradually moves to the center of the NW. As a consequence, volumetric expansion of the electrode is observed which induces stress in the electrode. Despite the NW electrode without a pre-existing crack \cite{eidelainsights}, here the volumetric expansion is not isotropic and the electrode expands more in the direction perpendicular to the crack length. At early stages of lithiation, when higher concentrations are confined to the outer region of the model, the corresponding volume expansion in this region is hindered by the interior domain which causes compressive hydrostatic stress in the outer region and tensile hydrostatic stress in the inner region of the NW. It is also observed that hydrostatic stress concentrates at the vicinity of the initial crack tips, then it reduces by time which makes the crack more stable. The above outcomes are in agreement with the previous studies in \cite{hu2017surface, zuo2016phase, chen2016analytical, wang2018three, gwak2018cohesive}. In addition, by the arrival of concentration front to the vicinity of the crack tips, the concentration distribution in electrode becomes nonuniform and a very high accumulation of $c$ is observed at the tips. This is mainly because that Li-ions transfer from lower hydrostatic stress regions towards higher hydrostatic stress regions based on \eqref{eq.massconservation}, which was reported in \cite{zuo2015phase, guan2018phase, wang2018peridynamic}. Contrary to the $c$ distribution, the distribution of hydrostatic stress becomes uniformed by time evolution. For comprehensive descriptions on the dynamics of lithiation in Si NW one can refer to \cite{eidelainsights}.

\begin{figure}[htbp]
	\begin{center}
		\begin{tabular}{c}
			\hspace{4em} {\large $t=0\,s$} \hspace{3.8em} {\large $t=2\,s$} \hspace{4.3em} {\large $t=3\,s$} \hspace{4.7em} {\large $t=5\,s$} \hspace{4.9em} {\large $t=6\,s$} \hspace{3em} \\
            \includegraphics[scale=0.21]{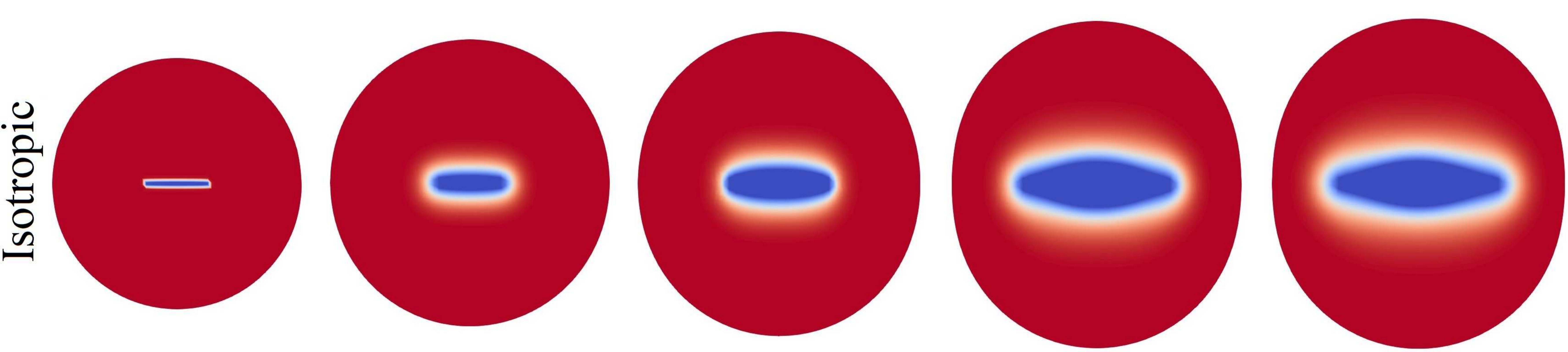} \\
            \includegraphics[scale=0.21]{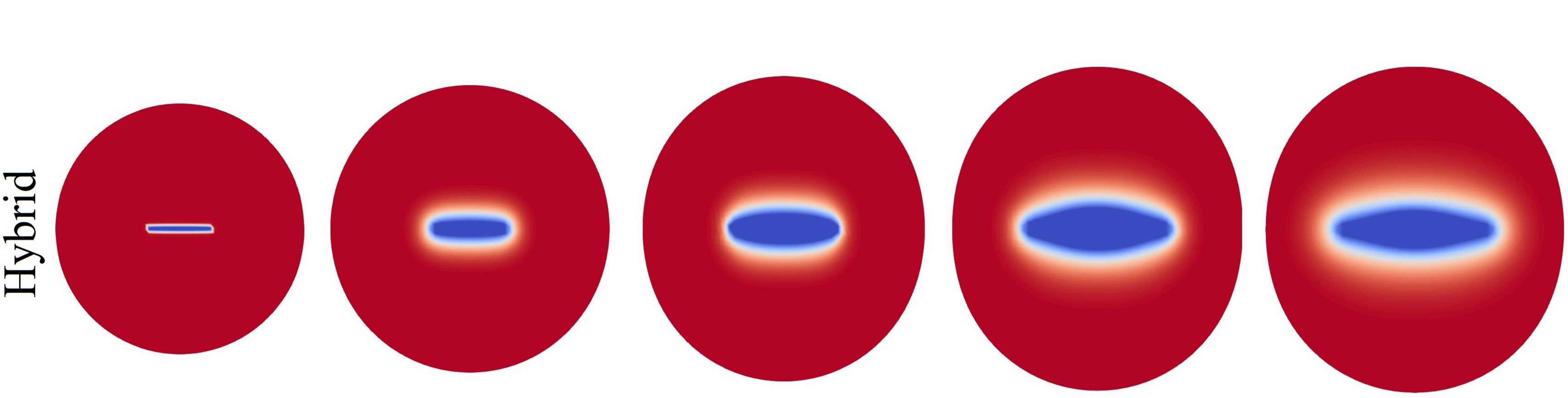} \\
            \includegraphics[scale=0.21]{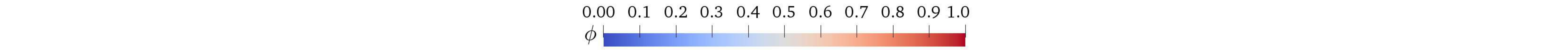} \\
		\end{tabular}
		\caption{Crack propagation for isotropic and hybrid models.} \label{fig.cont-phi}
	\end{center}
\end{figure}

Fig. \ref{fig.cont-c} does not disclose a remarkable distinction between isotropic model and hybrid model for the evolutions of $c$ and $g(\phi)\sigma_p$. However, as it was shown, the outer regions of the electrode undergo compressive hydrostatic stress. Therefore, it is expected that when the initial crack grows and the crack tips reach to the regions under compression, hybrid model shows less tendency to crack growth than the isotropic one. Fig. \ref{fig.cont-phi} shows the crack propagation during the charging process for isotropic and hybrid models. It is observed that at early stages of the process, both cracks start to propagate with a similar trend. But as predicted, at the final stages, crack growth in the hybrid model is impeded, unlike the isotropic model. Also, the grown crack in the isotropic model is evidently thicker, which indicates that the hybrid model has less crack growth tendency toward outer regions which are under compression. As mentioned before, the presence of a crack at the center of the electrode plate leads to anisotropic expansion. The length of the electrode in $x$ direction at $t=6$~s is $142.8$~nm for the isotropic model and $143.8$~nm for the hybrid model, and in $y$ direction is $161.1$~nm for the isotropic model and $158.8$~nm for the hybrid model. So, the measured values indicate that anisotropic expansion and hence crack growth are more evident in the isotropic model than the hybrid one.

\begin{figure}[H]
\begin{center}
\begin{tabular}{c c}
  \includegraphics[scale=0.38]{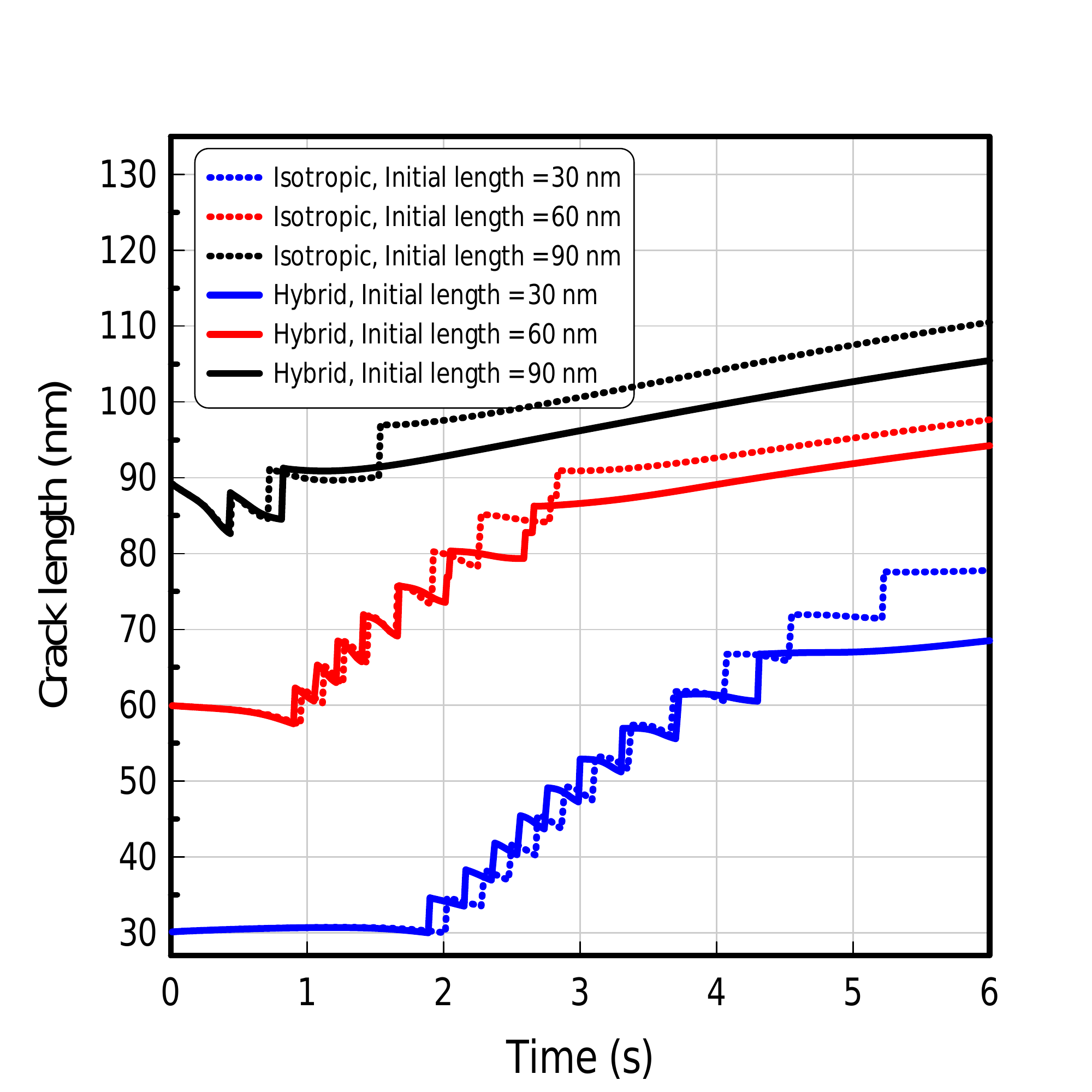} & \includegraphics[scale=0.38]{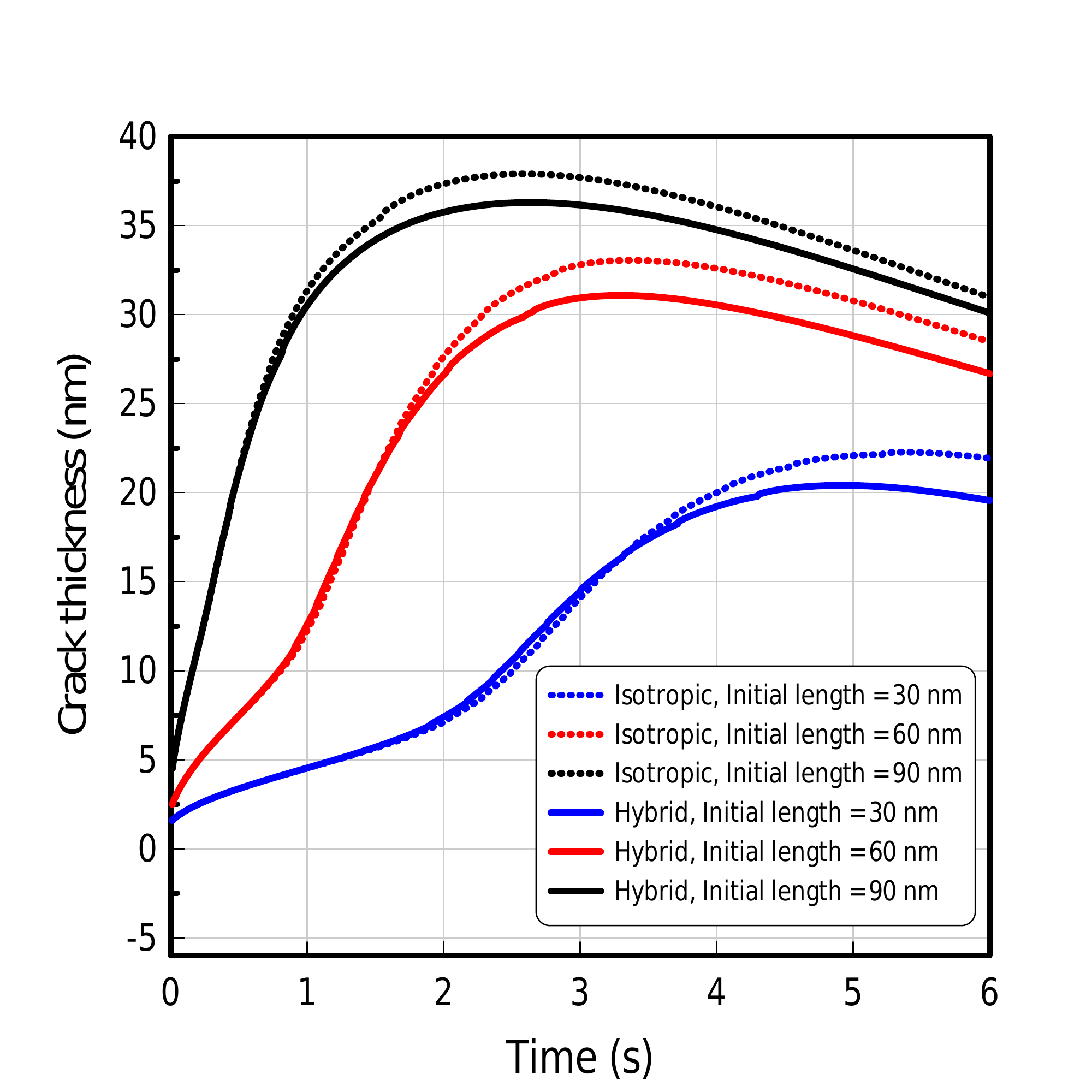} \\
  (a) & (b) \\
\end{tabular}
\caption{(a) Crack length and (b) crack thickness versus time for isotropic and hybrid models.}\label{fig.isohyb}
\end{center}
\end{figure}

Fig. \ref{fig.isohyb} compares the crack length and crack thickness of NWs with different initial crack lengths for isotropic and hybrid models over the time. Before the start of crack propagation, the crack is compacted due to anisotropic volumetric expansion which causes a small drop of the crack length. This phenomenon is more pronounced for NW electrodes with larger initial crack. In both models, cracks start to propagate with almost a constant slope until reaching certain values. Then the crack growth stops while crack length continues to increase slightly due to the volumetric expansion. The crack growth starts sooner and also ends sooner for the NWs with larger initial cracks. At the $t=6$~s, the crack lengths are increased $62.77\%$, $22.79\%$ and $159.30\%$ for isotropic models, and $57.05\%$, $17.18\%$ and $128.38\%$ for hybrid models, respectively for initial crack lengths of $30$~nm, $60$~nm and $90$~nm. Obviously, the hybrid models show smaller final crack length since they prevent crack growth in compressive regions. The crack lengths for isotropic models at $t=6$~s are $13.54\%$, $3.64\%$ and $4.78\%$ bigger than hybrid models for initial lengths of $30$~nm, $60$~nm and $90$~nm, respectively. Furthermore, Fig. \ref{fig.isohyb} (b) shows that cracks get thicker in the isotropic models which also was observed in Fig. \ref{fig.cont-phi}. It is also revealed that crack thickness increases faster in the models with longer initial cracks.

\subsection{Effect of fracture properties}\label{sec.NUM.Fracture}
\begin{figure}[htbp]
\begin{center}
\begin{tabular}{c c}
  \includegraphics[scale=0.38]{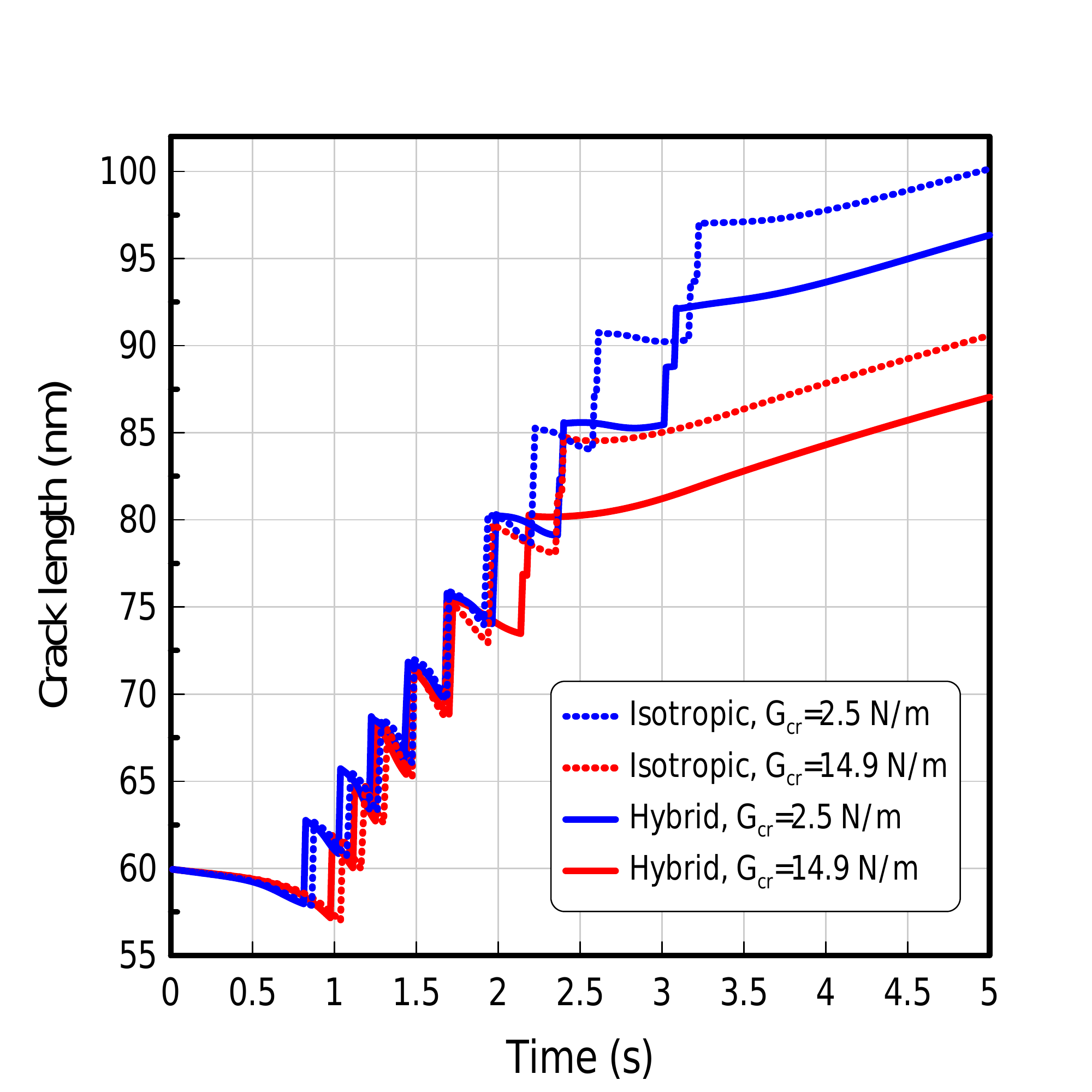} & \includegraphics[scale=0.38]{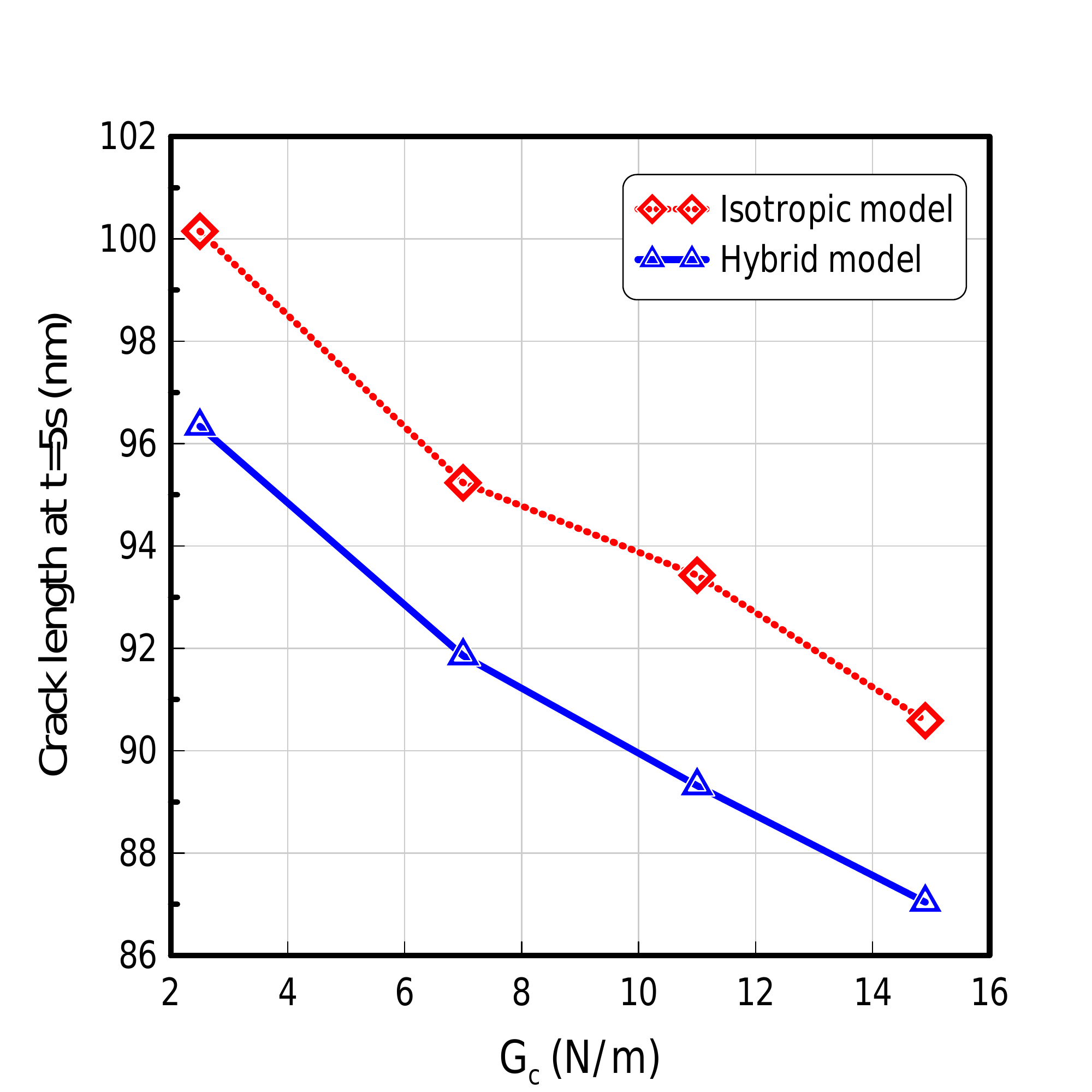} \\
  (a) & (b) \\
\end{tabular}
\caption{(a) Crack length versus time for different values of $G_{cr}$ and (b) Crack length at $t=5$~s versus $G_cr$.} \label{fig.gc}
\end{center}
\end{figure}

The crack length during the lithiation process for two different values of $G_{cr}$ are plotted over the time in Fig. \ref{fig.gc}~(a). Crack propagation in models with lesser critical energy release rate starts a little sooner and ends later with significantly bigger crack lengths. Furthermore, the effect of $G_{cr}$ on the crack length at $t=5$~s for isotropic and hybrid models is shown in Fig. \ref{fig.gc}~(b). The increase of the critical energy release rate reduces the crack length for both isotopic and hybrid models with a linear pattern, while, as expected, hybrid models show an average of $4.1\%$ less crack length.

\begin{figure}[htbp]
\begin{center}
\begin{tabular}{c c}
    \includegraphics[scale=0.38]{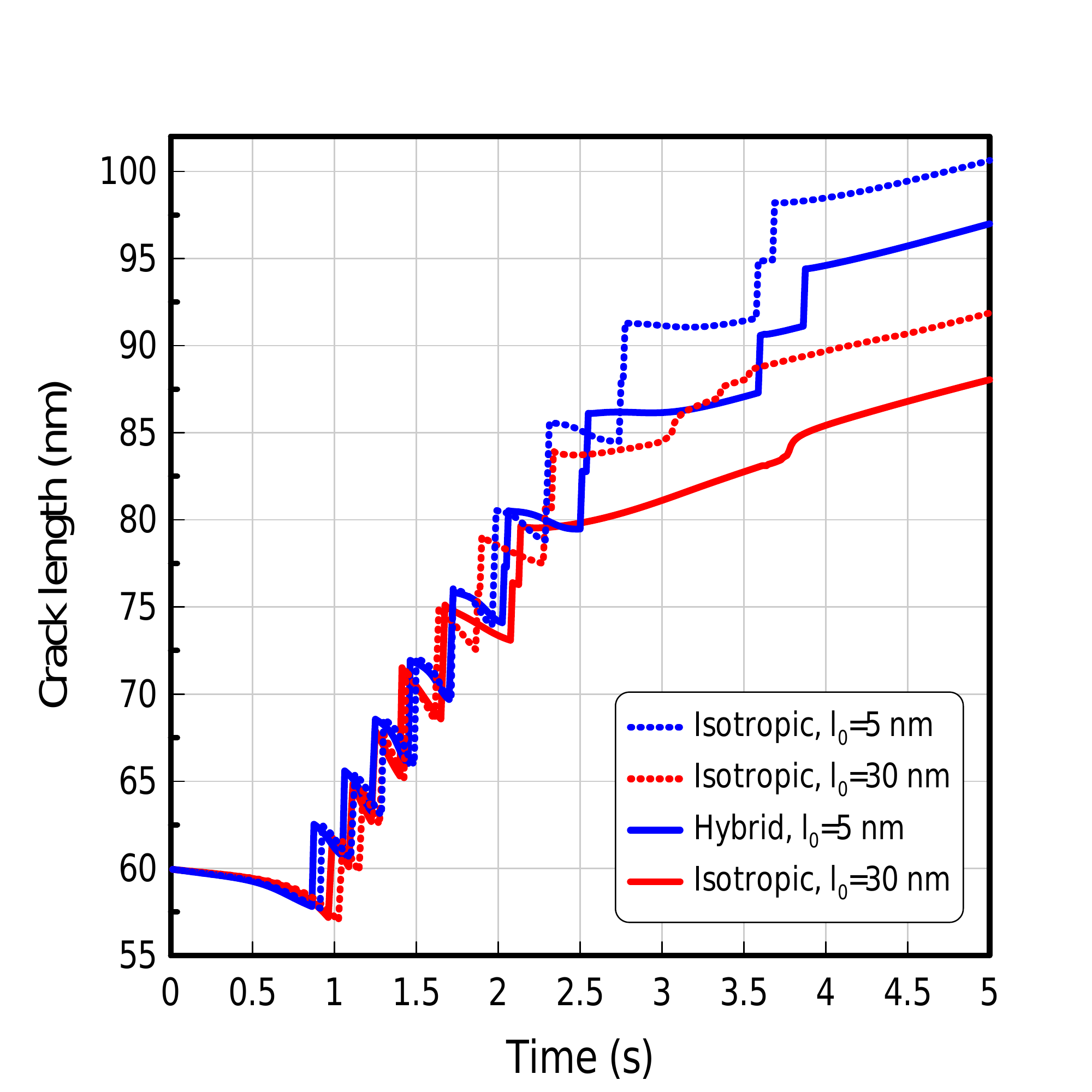} & \includegraphics[scale=0.38]{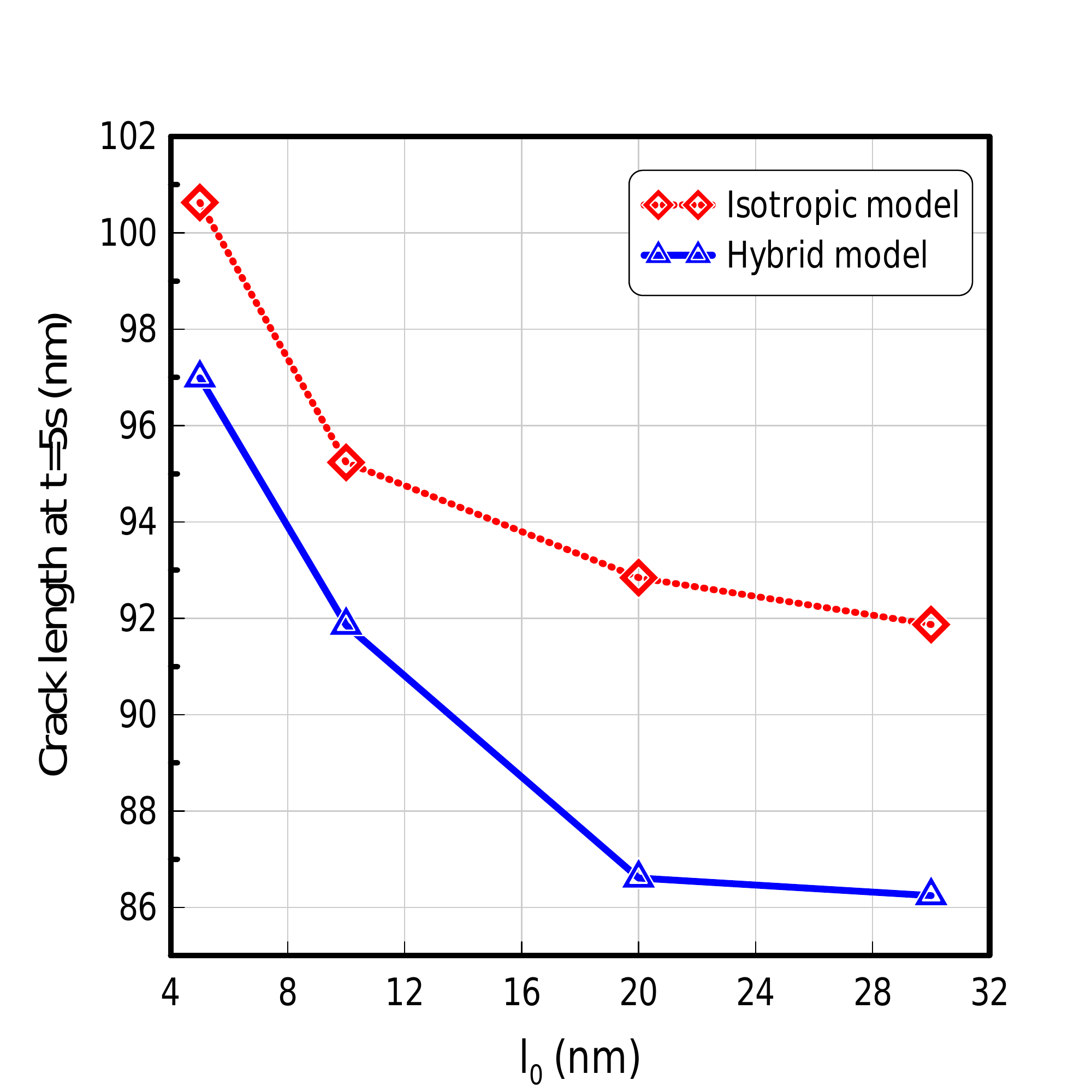} \\
    (a) & (b) \\
\end{tabular}
\caption{(a) Crack length versus time for different values of $l_0$ and (b) Crack length at $t=5$~s versus $l_0$.} \label{fig.l0}
\end{center}
\end{figure}

Fig. \ref{fig.l0}~(a) illustrates the evolution of crack length versus time for different values of $l_0$. Similar to the critical energy release rate, crack propagation for the models with lesser regularization constant starts sooner and ends later with bigger crack lengths. Also, the increase of $l_0$ reduces the crack length for both isotopic and hybrid models with an exponential pattern, as shown in Fig. \ref{fig.l0}~(b). The isotropic models show $3.74\%$, $3.69\%$, $7.19\%$ and $6.53\%$ larger crack length than the hybrid models for $l_0=5$~nm, $l_0=10$~nm, $l_0=20$~nm and $l_0=30$~nm, respectively.

\section{Conclusions}\label{sec.Conclusions}
In this contribution, a hybrid phase field model was proposed to investigate the fracture process of electrodes in Li-ion batteries. The framework links Li-ion diffusion, stress evolution and crack propagation. The model takes account of the dependency of Elastic properties on the Li concentration. The variational approach along with the finite element method was conducted to obtain the discretized governing equations of the system on space domain. The model was implemented in MATLAB, and all simulations were performed in the same software. Unlike the traditional isotopic models, the proposed hybrid formulation aims to prevent crack growth in the region under compression. Both isotopic and hybrid models were applied to a silicon nanowire electrode particle, and the effects of different initial crack lengths and fracture properties were investigated. The numerical results shown that the hybrid model shows less tendency to crack growth than the isotropic model. Moreover, the increase of the critical energy release rate and regularization constant reduces the crack length with linear and exponential patterns, respectively. The proposed hybrid model has the potential to assist the understanding of the fracture behavior of Li-ion battery electrodes and provides insightful guidelines for the design of next-generation electrodes in the future. It should be noted that the present work is only a preliminary study, and to demonstrate the capability and potential applicability of the proposed hybrid approach there are still different detailed aspects that must be explored. For example, electrodes with geometrical constraints which undergo huge compressive stresses \cite{zhang2016variational, lee2015kinetics}, delithiation process of an electrode, in which compressive stress is generated in the inner part of the electrode \cite{xu2016phase, klinsmann2016modeling2, klinsmann2016modeling, liu2011reversible}, and electrode with multiple pre-existing cracks \cite{zuo2015phase, wang2018peridynamic, xu2016phase, wang2018three, miehe2016phase, klinsmann2016modeling2}. Besides, different electrode particles with different shapes and sizes must be investigated.

\bigskip

{\bf Declarations of interest:} None.

\bibliography{mybibfile}

\begin{thebibliography}{62}
\expandafter\ifx\csname natexlab\endcsname\relax\def\natexlab#1{#1}\fi
\providecommand{\url}[1]{\texttt{#1}}
\providecommand{\href}[2]{#2}
\providecommand{\path}[1]{#1}
\providecommand{\DOIprefix}{doi:}
\providecommand{\ArXivprefix}{arXiv:}
\providecommand{\URLprefix}{URL: }
\providecommand{\Pubmedprefix}{pmid:}
\providecommand{\doi}[1]{\href{http://dx.doi.org/#1}{\path{#1}}}
\providecommand{\Pubmed}[1]{\href{pmid:#1}{\path{#1}}}
\providecommand{\bibinfo}[2]{#2}
\ifx\xfnm\relax \def\xfnm[#1]{\unskip,\space#1}\fi
%Type = Article
\bibitem[{Smith(2010)}]{smith2010electrochemical}
\bibinfo{author}{K.~A. Smith},
\newblock \bibinfo{title}{Electrochemical control of lithium-ion batteries},
\newblock \bibinfo{journal}{IEEE Control systems magazine} \bibinfo{volume}{30}
  (\bibinfo{year}{2010}) \bibinfo{pages}{18--25}.
%Type = Article
\bibitem[{Zhao et~al.(2019)Zhao, Stein, Bai, Al-Siraj, Yang, and
  Xu}]{zhao2019review}
\bibinfo{author}{Y.~Zhao}, \bibinfo{author}{P.~Stein},
  \bibinfo{author}{Y.~Bai}, \bibinfo{author}{M.~Al-Siraj},
  \bibinfo{author}{Y.~Yang}, \bibinfo{author}{B.-X. Xu},
\newblock \bibinfo{title}{A review on modeling of electro-chemo-mechanics in
  lithium-ion batteries},
\newblock \bibinfo{journal}{Journal of Power Sources} \bibinfo{volume}{413}
  (\bibinfo{year}{2019}) \bibinfo{pages}{259--283}.
%Type = Article
\bibitem[{Mukhopadhyay and Sheldon(2014)}]{mukhopadhyay2014deformation}
\bibinfo{author}{A.~Mukhopadhyay}, \bibinfo{author}{B.~W. Sheldon},
\newblock \bibinfo{title}{Deformation and stress in electrode materials for
  li-ion batteries},
\newblock \bibinfo{journal}{Progress in Materials Science} \bibinfo{volume}{63}
  (\bibinfo{year}{2014}) \bibinfo{pages}{58--116}.
%Type = Article
\bibitem[{Kumar et~al.(2018)Kumar, Nehra, Kedia, Dilbaghi, Tankeshwar, and
  Kim}]{kumar2018carbon}
\bibinfo{author}{S.~Kumar}, \bibinfo{author}{M.~Nehra},
  \bibinfo{author}{D.~Kedia}, \bibinfo{author}{N.~Dilbaghi},
  \bibinfo{author}{K.~Tankeshwar}, \bibinfo{author}{K.-H. Kim},
\newblock \bibinfo{title}{Carbon nanotubes: A potential material for energy
  conversion and storage},
\newblock \bibinfo{journal}{Progress in energy and combustion science}
  \bibinfo{volume}{64} (\bibinfo{year}{2018}) \bibinfo{pages}{219--253}.
%Type = Article
\bibitem[{Obrovac and Christensen(2004)}]{obrovac2004structural}
\bibinfo{author}{M.~Obrovac}, \bibinfo{author}{L.~Christensen},
\newblock \bibinfo{title}{Structural changes in silicon anodes during lithium
  insertion/extraction},
\newblock \bibinfo{journal}{Electrochemical and solid-state letters}
  \bibinfo{volume}{7} (\bibinfo{year}{2004}) \bibinfo{pages}{A93--A96}.
%Type = Article
\bibitem[{Ebner et~al.(2013)Ebner, Marone, Stampanoni, and
  Wood}]{ebner2013visualization}
\bibinfo{author}{M.~Ebner}, \bibinfo{author}{F.~Marone},
  \bibinfo{author}{M.~Stampanoni}, \bibinfo{author}{V.~Wood},
\newblock \bibinfo{title}{Visualization and quantification of electrochemical
  and mechanical degradation in li ion batteries},
\newblock \bibinfo{journal}{Science} \bibinfo{volume}{342}
  (\bibinfo{year}{2013}) \bibinfo{pages}{716--720}.
%Type = Article
\bibitem[{Zuo and Zhao(2015)}]{zuo2015phase}
\bibinfo{author}{P.~Zuo}, \bibinfo{author}{Y.-P. Zhao},
\newblock \bibinfo{title}{A phase field model coupling lithium diffusion and
  stress evolution with crack propagation and application in lithium ion
  batteries},
\newblock \bibinfo{journal}{Physical chemistry chemical physics}
  \bibinfo{volume}{17} (\bibinfo{year}{2015}) \bibinfo{pages}{287--297}.
%Type = Article
\bibitem[{Ryu et~al.(2011)Ryu, Choi, Cui, and Nix}]{ryu2011size}
\bibinfo{author}{I.~Ryu}, \bibinfo{author}{J.~W. Choi},
  \bibinfo{author}{Y.~Cui}, \bibinfo{author}{W.~D. Nix},
\newblock \bibinfo{title}{Size-dependent fracture of si nanowire battery
  anodes},
\newblock \bibinfo{journal}{Journal of the mechanics and physics of solids}
  \bibinfo{volume}{59} (\bibinfo{year}{2011}) \bibinfo{pages}{1717--1730}.
%Type = Article
\bibitem[{Grantab and Shenoy(2012)}]{grantab2012pressure}
\bibinfo{author}{R.~Grantab}, \bibinfo{author}{V.~B. Shenoy},
\newblock \bibinfo{title}{Pressure-gradient dependent diffusion and crack
  propagation in lithiated silicon nanowires},
\newblock \bibinfo{journal}{Journal of the Electrochemical Society}
  \bibinfo{volume}{159} (\bibinfo{year}{2012}) \bibinfo{pages}{A584--A591}.
%Type = Article
\bibitem[{Chen et~al.(2016)Chen, Zhou, and Cai}]{chen2016analytical}
\bibinfo{author}{B.~Chen}, \bibinfo{author}{J.~Zhou}, \bibinfo{author}{R.~Cai},
\newblock \bibinfo{title}{Analytical model for crack propagation in spherical
  nano electrodes of lithium-ion batteries},
\newblock \bibinfo{journal}{Electrochimica Acta} \bibinfo{volume}{210}
  (\bibinfo{year}{2016}) \bibinfo{pages}{7--14}.
%Type = Article
\bibitem[{Hu et~al.(2017)Hu, Zhao, Cai, and Zhou}]{hu2017surface}
\bibinfo{author}{X.~Hu}, \bibinfo{author}{Y.~Zhao}, \bibinfo{author}{R.~Cai},
  \bibinfo{author}{J.~Zhou},
\newblock \bibinfo{title}{Surface effected fracture behavior of nano-spherical
  electrodes during lithiation reaction},
\newblock \bibinfo{journal}{Materials Science and Engineering: A}
  \bibinfo{volume}{707} (\bibinfo{year}{2017}) \bibinfo{pages}{92--100}.
%Type = Article
\bibitem[{Wang et~al.(2018{\natexlab{a}})Wang, Oterkus, and
  Oterkus}]{wang2018predicting}
\bibinfo{author}{H.~Wang}, \bibinfo{author}{E.~Oterkus},
  \bibinfo{author}{S.~Oterkus},
\newblock \bibinfo{title}{Predicting fracture evolution during lithiation
  process using peridynamics},
\newblock \bibinfo{journal}{Engineering Fracture Mechanics}
  \bibinfo{volume}{192} (\bibinfo{year}{2018}{\natexlab{a}})
  \bibinfo{pages}{176--191}.
%Type = Article
\bibitem[{Wang et~al.(2018{\natexlab{b}})Wang, Oterkus, and
  Oterkus}]{wang2018peridynamic}
\bibinfo{author}{H.~Wang}, \bibinfo{author}{E.~Oterkus},
  \bibinfo{author}{S.~Oterkus},
\newblock \bibinfo{title}{Peridynamic modelling of fracture in marine
  lithium-ion batteries},
\newblock \bibinfo{journal}{Ocean Engineering} \bibinfo{volume}{151}
  (\bibinfo{year}{2018}{\natexlab{b}}) \bibinfo{pages}{257--267}.
%Type = Article
\bibitem[{Wang et~al.(2018{\natexlab{c}})Wang, Oterkus, and
  Oterkus}]{wang2018three}
\bibinfo{author}{H.~Wang}, \bibinfo{author}{E.~Oterkus},
  \bibinfo{author}{S.~Oterkus},
\newblock \bibinfo{title}{Three-dimensional peridynamic model for predicting
  fracture evolution during the lithiation process},
\newblock \bibinfo{journal}{Energies} \bibinfo{volume}{11}
  (\bibinfo{year}{2018}{\natexlab{c}}) \bibinfo{pages}{1461}.
%Type = Article
\bibitem[{Gwak et~al.(2018)Gwak, Jin, and Cho}]{gwak2018cohesive}
\bibinfo{author}{Y.~Gwak}, \bibinfo{author}{Y.~Jin}, \bibinfo{author}{M.~Cho},
\newblock \bibinfo{title}{Cohesive zone model for crack propagation in
  crystalline silicon nanowires},
\newblock \bibinfo{journal}{Journal of mechanical science and technology}
  \bibinfo{volume}{32} (\bibinfo{year}{2018}) \bibinfo{pages}{3755--3763}.
%Type = Article
\bibitem[{Hu et~al.(2010)Hu, Zhao, and Suo}]{hu2010averting}
\bibinfo{author}{Y.~Hu}, \bibinfo{author}{X.~Zhao}, \bibinfo{author}{Z.~Suo},
\newblock \bibinfo{title}{Averting cracks caused by insertion reaction in
  lithium--ion batteries},
\newblock \bibinfo{journal}{Journal of Materials Research} \bibinfo{volume}{25}
  (\bibinfo{year}{2010}) \bibinfo{pages}{1007--1010}.
%Type = Article
\bibitem[{Steinbach(2009)}]{steinbach2009phase}
\bibinfo{author}{I.~Steinbach},
\newblock \bibinfo{title}{Phase-field models in materials science},
\newblock \bibinfo{journal}{Modelling and simulation in materials science and
  engineering} \bibinfo{volume}{17} (\bibinfo{year}{2009})
  \bibinfo{pages}{073001}.
%Type = Article
\bibitem[{Bourdin et~al.(2000)Bourdin, Francfort, and
  Marigo}]{bourdin2000numerical}
\bibinfo{author}{B.~Bourdin}, \bibinfo{author}{G.~A. Francfort},
  \bibinfo{author}{J.-J. Marigo},
\newblock \bibinfo{title}{Numerical experiments in revisited brittle fracture},
\newblock \bibinfo{journal}{Journal of the Mechanics and Physics of Solids}
  \bibinfo{volume}{48} (\bibinfo{year}{2000}) \bibinfo{pages}{797--826}.
%Type = Article
\bibitem[{Bourdin et~al.(2008)Bourdin, Francfort, and
  Marigo}]{bourdin2008variational}
\bibinfo{author}{B.~Bourdin}, \bibinfo{author}{G.~A. Francfort},
  \bibinfo{author}{J.-J. Marigo},
\newblock \bibinfo{title}{The variational approach to fracture},
\newblock \bibinfo{journal}{Journal of elasticity} \bibinfo{volume}{91}
  (\bibinfo{year}{2008}) \bibinfo{pages}{5--148}.
%Type = Article
\bibitem[{Aranson et~al.(2000)Aranson, Kalatsky, and
  Vinokur}]{aranson2000continuum}
\bibinfo{author}{I.~Aranson}, \bibinfo{author}{V.~Kalatsky},
  \bibinfo{author}{V.~Vinokur},
\newblock \bibinfo{title}{Continuum field description of crack propagation},
\newblock \bibinfo{journal}{Physical review letters} \bibinfo{volume}{85}
  (\bibinfo{year}{2000}) \bibinfo{pages}{118}.
%Type = Article
\bibitem[{Karma et~al.(2001)Karma, Kessler, and Levine}]{karma2001phase}
\bibinfo{author}{A.~Karma}, \bibinfo{author}{D.~A. Kessler},
  \bibinfo{author}{H.~Levine},
\newblock \bibinfo{title}{Phase-field model of mode iii dynamic fracture},
\newblock \bibinfo{journal}{Physical Review Letters} \bibinfo{volume}{87}
  (\bibinfo{year}{2001}) \bibinfo{pages}{045501}.
%Type = Article
\bibitem[{Hakim and Karma(2009)}]{hakim2009laws}
\bibinfo{author}{V.~Hakim}, \bibinfo{author}{A.~Karma},
\newblock \bibinfo{title}{Laws of crack motion and phase-field models of
  fracture},
\newblock \bibinfo{journal}{Journal of the Mechanics and Physics of Solids}
  \bibinfo{volume}{57} (\bibinfo{year}{2009}) \bibinfo{pages}{342--368}.
%Type = Article
\bibitem[{Spatschek et~al.(2011)Spatschek, Brener, and
  Karma}]{spatschek2011phase}
\bibinfo{author}{R.~Spatschek}, \bibinfo{author}{E.~Brener},
  \bibinfo{author}{A.~Karma},
\newblock \bibinfo{title}{Phase field modeling of crack propagation},
\newblock \bibinfo{journal}{Philosophical Magazine} \bibinfo{volume}{91}
  (\bibinfo{year}{2011}) \bibinfo{pages}{75--95}.
%Type = Article
\bibitem[{Lu et~al.(2019)Lu, Li, Tie, Hou, and Zhang}]{lu2019crack}
\bibinfo{author}{X.~Lu}, \bibinfo{author}{C.~Li}, \bibinfo{author}{Y.~Tie},
  \bibinfo{author}{Y.~Hou}, \bibinfo{author}{C.~Zhang},
\newblock \bibinfo{title}{Crack propagation simulation in brittle elastic
  materials by a phase field method},
\newblock \bibinfo{journal}{Theoretical and Applied Mechanics Letters}
  \bibinfo{volume}{9} (\bibinfo{year}{2019}) \bibinfo{pages}{339--352}.
%Type = Article
\bibitem[{Miehe et~al.(2016)Miehe, Dal, Sch{\"a}nzel, and
  Raina}]{miehe2016phase}
\bibinfo{author}{C.~Miehe}, \bibinfo{author}{H.~Dal}, \bibinfo{author}{L.-M.
  Sch{\"a}nzel}, \bibinfo{author}{A.~Raina},
\newblock \bibinfo{title}{A phase-field model for chemo-mechanical induced
  fracture in lithium-ion battery electrode particles},
\newblock \bibinfo{journal}{International journal for numerical methods in
  engineering} \bibinfo{volume}{106} (\bibinfo{year}{2016})
  \bibinfo{pages}{683--711}.
%Type = Article
\bibitem[{Zuo and Zhao(2016)}]{zuo2016phase}
\bibinfo{author}{P.~Zuo}, \bibinfo{author}{Y.-P. Zhao},
\newblock \bibinfo{title}{Phase field modeling of lithium diffusion, finite
  deformation, stress evolution and crack propagation in lithium ion battery},
\newblock \bibinfo{journal}{Extreme mechanics letters} \bibinfo{volume}{9}
  (\bibinfo{year}{2016}) \bibinfo{pages}{467--479}.
%Type = Article
\bibitem[{Guan et~al.(2018)Guan, Liu, and Gao}]{guan2018phase}
\bibinfo{author}{P.~Guan}, \bibinfo{author}{L.~Liu}, \bibinfo{author}{Y.~Gao},
\newblock \bibinfo{title}{Phase-field modeling of solid electrolyte interphase
  (sei) cracking in lithium batteries},
\newblock \bibinfo{journal}{ECS Transactions} \bibinfo{volume}{85}
  (\bibinfo{year}{2018}) \bibinfo{pages}{1041--1051}.
%Type = Article
\bibitem[{Amor et~al.(2009)Amor, Marigo, and Maurini}]{amor2009regularized}
\bibinfo{author}{H.~Amor}, \bibinfo{author}{J.-J. Marigo},
  \bibinfo{author}{C.~Maurini},
\newblock \bibinfo{title}{Regularized formulation of the variational brittle
  fracture with unilateral contact: Numerical experiments},
\newblock \bibinfo{journal}{Journal of the Mechanics and Physics of Solids}
  \bibinfo{volume}{57} (\bibinfo{year}{2009}) \bibinfo{pages}{1209--1229}.
%Type = Article
\bibitem[{Miehe et~al.(2010)Miehe, Welschinger, and
  Hofacker}]{miehe2010thermodynamically}
\bibinfo{author}{C.~Miehe}, \bibinfo{author}{F.~Welschinger},
  \bibinfo{author}{M.~Hofacker},
\newblock \bibinfo{title}{Thermodynamically consistent phase-field models of
  fracture: Variational principles and multi-field fe implementations},
\newblock \bibinfo{journal}{International Journal for Numerical Methods in
  Engineering} \bibinfo{volume}{83} (\bibinfo{year}{2010})
  \bibinfo{pages}{1273--1311}.
%Type = Article
\bibitem[{Zhao et~al.(2016)Zhao, Xu, Stein, and Gross}]{zhao2016phase}
\bibinfo{author}{Y.~Zhao}, \bibinfo{author}{B.-X. Xu},
  \bibinfo{author}{P.~Stein}, \bibinfo{author}{D.~Gross},
\newblock \bibinfo{title}{Phase-field study of electrochemical reactions at
  exterior and interior interfaces in li-ion battery electrode particles},
\newblock \bibinfo{journal}{Computer methods in applied mechanics and
  engineering} \bibinfo{volume}{312} (\bibinfo{year}{2016})
  \bibinfo{pages}{428--446}.
%Type = Article
\bibitem[{Zhang et~al.(2016)Zhang, Krischok, and Linder}]{zhang2016variational}
\bibinfo{author}{X.~Zhang}, \bibinfo{author}{A.~Krischok},
  \bibinfo{author}{C.~Linder},
\newblock \bibinfo{title}{A variational framework to model diffusion induced
  large plastic deformation and phase field fracture during initial two-phase
  lithiation of silicon electrodes},
\newblock \bibinfo{journal}{Computer methods in applied mechanics and
  engineering} \bibinfo{volume}{312} (\bibinfo{year}{2016})
  \bibinfo{pages}{51--77}.
%Type = Article
\bibitem[{Klinsmann et~al.(2016{\natexlab{a}})Klinsmann, Rosato, Kamlah, and
  McMeeking}]{klinsmann2016modeling2}
\bibinfo{author}{M.~Klinsmann}, \bibinfo{author}{D.~Rosato},
  \bibinfo{author}{M.~Kamlah}, \bibinfo{author}{R.~M. McMeeking},
\newblock \bibinfo{title}{Modeling crack growth during li extraction in storage
  particles using a fracture phase field approach},
\newblock \bibinfo{journal}{Journal of the electrochemical society}
  \bibinfo{volume}{163} (\bibinfo{year}{2016}{\natexlab{a}})
  \bibinfo{pages}{A102--A118}.
%Type = Article
\bibitem[{Klinsmann et~al.(2016{\natexlab{b}})Klinsmann, Rosato, Kamlah, and
  McMeeking}]{klinsmann2016modeling3}
\bibinfo{author}{M.~Klinsmann}, \bibinfo{author}{D.~Rosato},
  \bibinfo{author}{M.~Kamlah}, \bibinfo{author}{R.~M. McMeeking},
\newblock \bibinfo{title}{Modeling crack growth during li insertion in storage
  particles using a fracture phase field approach},
\newblock \bibinfo{journal}{Journal of the Mechanics and Physics of Solids}
  \bibinfo{volume}{92} (\bibinfo{year}{2016}{\natexlab{b}})
  \bibinfo{pages}{313--344}.
%Type = Article
\bibitem[{Klinsmann et~al.(2016{\natexlab{c}})Klinsmann, Rosato, Kamlah, and
  McMeeking}]{klinsmann2016modeling}
\bibinfo{author}{M.~Klinsmann}, \bibinfo{author}{D.~Rosato},
  \bibinfo{author}{M.~Kamlah}, \bibinfo{author}{R.~M. McMeeking},
\newblock \bibinfo{title}{Modeling crack growth during li extraction and
  insertion within the second half cycle},
\newblock \bibinfo{journal}{Journal of power sources} \bibinfo{volume}{331}
  (\bibinfo{year}{2016}{\natexlab{c}}) \bibinfo{pages}{32--42}.
%Type = Article
\bibitem[{Xu et~al.(2016)Xu, Zhao, and Stein}]{xu2016phase}
\bibinfo{author}{B.-X. Xu}, \bibinfo{author}{Y.~Zhao},
  \bibinfo{author}{P.~Stein},
\newblock \bibinfo{title}{Phase field modeling of electrochemically induced
  fracture in li-ion battery with large deformation and phase segregation},
\newblock \bibinfo{journal}{GAMM-Mitteilungen} \bibinfo{volume}{39}
  (\bibinfo{year}{2016}) \bibinfo{pages}{92--109}.
%Type = Article
\bibitem[{Nguyen et~al.(2018)Nguyen, Bolivar, Shi, R{\'e}thor{\'e}, King,
  Fregonese, Adrien, Buffiere, and Baietto}]{nguyen2018phase}
\bibinfo{author}{T.~T. Nguyen}, \bibinfo{author}{J.~Bolivar},
  \bibinfo{author}{Y.~Shi}, \bibinfo{author}{J.~R{\'e}thor{\'e}},
  \bibinfo{author}{A.~King}, \bibinfo{author}{M.~Fregonese},
  \bibinfo{author}{J.~Adrien}, \bibinfo{author}{J.-Y. Buffiere},
  \bibinfo{author}{M.-C. Baietto},
\newblock \bibinfo{title}{A phase field method for modeling anodic dissolution
  induced stress corrosion crack propagation},
\newblock \bibinfo{journal}{Corrosion Science} \bibinfo{volume}{132}
  (\bibinfo{year}{2018}) \bibinfo{pages}{146--160}.
%Type = Article
\bibitem[{Ambati et~al.(2015)Ambati, Gerasimov, and
  De~Lorenzis}]{ambati2015review}
\bibinfo{author}{M.~Ambati}, \bibinfo{author}{T.~Gerasimov},
  \bibinfo{author}{L.~De~Lorenzis},
\newblock \bibinfo{title}{A review on phase-field models of brittle fracture
  and a new fast hybrid formulation},
\newblock \bibinfo{journal}{Computational Mechanics} \bibinfo{volume}{55}
  (\bibinfo{year}{2015}) \bibinfo{pages}{383--405}.
%Type = Article
\bibitem[{Wu(2017)}]{wu2017unified}
\bibinfo{author}{J.-Y. Wu},
\newblock \bibinfo{title}{A unified phase-field theory for the mechanics of
  damage and quasi-brittle failure},
\newblock \bibinfo{journal}{Journal of the Mechanics and Physics of Solids}
  \bibinfo{volume}{103} (\bibinfo{year}{2017}) \bibinfo{pages}{72--99}.
%Type = Article
\bibitem[{Wu(2018)}]{wu2018geometrically}
\bibinfo{author}{J.-Y. Wu},
\newblock \bibinfo{title}{A geometrically regularized gradient-damage model
  with energetic equivalence},
\newblock \bibinfo{journal}{Computer Methods in Applied Mechanics and
  Engineering} \bibinfo{volume}{328} (\bibinfo{year}{2018})
  \bibinfo{pages}{612--637}.
%Type = Article
\bibitem[{Wu and Nguyen(2018)}]{wu2018length}
\bibinfo{author}{J.-Y. Wu}, \bibinfo{author}{V.~P. Nguyen},
\newblock \bibinfo{title}{A length scale insensitive phase-field damage model
  for brittle fracture},
\newblock \bibinfo{journal}{Journal of the Mechanics and Physics of Solids}
  \bibinfo{volume}{119} (\bibinfo{year}{2018}) \bibinfo{pages}{20--42}.
%Type = Article
\bibitem[{Kuhn et~al.(2015)Kuhn, Schl{\"u}ter, and
  M{\"u}ller}]{kuhn2015degradation}
\bibinfo{author}{C.~Kuhn}, \bibinfo{author}{A.~Schl{\"u}ter},
  \bibinfo{author}{R.~M{\"u}ller},
\newblock \bibinfo{title}{On degradation functions in phase field fracture
  models},
\newblock \bibinfo{journal}{Computational Materials Science}
  \bibinfo{volume}{108} (\bibinfo{year}{2015}) \bibinfo{pages}{374--384}.
%Type = Article
\bibitem[{Kuhn and M{\"u}ller(2010)}]{kuhn2010continuum}
\bibinfo{author}{C.~Kuhn}, \bibinfo{author}{R.~M{\"u}ller},
\newblock \bibinfo{title}{A continuum phase field model for fracture},
\newblock \bibinfo{journal}{Engineering Fracture Mechanics}
  \bibinfo{volume}{77} (\bibinfo{year}{2010}) \bibinfo{pages}{3625--3634}.
%Type = Article
\bibitem[{Prussin(1961)}]{prussin1961generation}
\bibinfo{author}{S.~Prussin},
\newblock \bibinfo{title}{Generation and distribution of dislocations by solute
  diffusion},
\newblock \bibinfo{journal}{Journal of applied physics} \bibinfo{volume}{32}
  (\bibinfo{year}{1961}) \bibinfo{pages}{1876--1881}.
%Type = Article
\bibitem[{Li(1978)}]{li1978physical}
\bibinfo{author}{J.~C.-M. Li},
\newblock \bibinfo{title}{Physical chemistry of some microstructural
  phenomena},
\newblock \bibinfo{journal}{Metallurgical transactions A} \bibinfo{volume}{9}
  (\bibinfo{year}{1978}) \bibinfo{pages}{1353--1380}.
%Type = Article
\bibitem[{Singh et~al.(2008)Singh, Ceder, and Bazant}]{singh2008intercalation}
\bibinfo{author}{G.~K. Singh}, \bibinfo{author}{G.~Ceder},
  \bibinfo{author}{M.~Z. Bazant},
\newblock \bibinfo{title}{Intercalation dynamics in rechargeable battery
  materials: General theory and phase-transformation waves in lifepo4},
\newblock \bibinfo{journal}{Electrochimica acta} \bibinfo{volume}{53}
  (\bibinfo{year}{2008}) \bibinfo{pages}{7599--7613}.
%Type = Article
\bibitem[{Yang(2005)}]{yang2005interaction}
\bibinfo{author}{F.~Yang},
\newblock \bibinfo{title}{Interaction between diffusion and chemical stresses},
\newblock \bibinfo{journal}{Materials science and engineering: A}
  \bibinfo{volume}{409} (\bibinfo{year}{2005}) \bibinfo{pages}{153--159}.
%Type = Article
\bibitem[{Miehe et~al.(2010)Miehe, Hofacker, and Welschinger}]{miehe2010phase}
\bibinfo{author}{C.~Miehe}, \bibinfo{author}{M.~Hofacker},
  \bibinfo{author}{F.~Welschinger},
\newblock \bibinfo{title}{A phase field model for rate-independent crack
  propagation: Robust algorithmic implementation based on operator splits},
\newblock \bibinfo{journal}{Computer Methods in Applied Mechanics and
  Engineering} \bibinfo{volume}{199} (\bibinfo{year}{2010})
  \bibinfo{pages}{2765--2778}.
%Type = Article
\bibitem[{Chen(2002)}]{chen2002phase}
\bibinfo{author}{L.-Q. Chen},
\newblock \bibinfo{title}{Phase-field models for microstructure evolution},
\newblock \bibinfo{journal}{Annual review of materials research}
  \bibinfo{volume}{32} (\bibinfo{year}{2002}) \bibinfo{pages}{113--140}.
%Type = Article
\bibitem[{Xie et~al.(2015)Xie, Qiu, Gao, Guan, and Yuan}]{xie2015phase}
\bibinfo{author}{Y.~Xie}, \bibinfo{author}{M.~Qiu}, \bibinfo{author}{X.~Gao},
  \bibinfo{author}{D.~Guan}, \bibinfo{author}{C.~Yuan},
\newblock \bibinfo{title}{Phase field modeling of silicon nanowire based
  lithium ion battery composite electrode},
\newblock \bibinfo{journal}{Electrochimica acta} \bibinfo{volume}{186}
  (\bibinfo{year}{2015}) \bibinfo{pages}{542--551}.
%Type = Article
\bibitem[{Wu(2018)}]{wu2018robust}
\bibinfo{author}{J.-Y. Wu},
\newblock \bibinfo{title}{Robust numerical implementation of non-standard
  phase-field damage models for failure in solids},
\newblock \bibinfo{journal}{Computer Methods in Applied Mechanics and
  Engineering} \bibinfo{volume}{340} (\bibinfo{year}{2018})
  \bibinfo{pages}{767--797}.
%Type = Inproceedings
\bibitem[{Newmark et~al.(1959)}]{newmark1959method}
\bibinfo{author}{N.~M. Newmark}, et~al.,
\newblock \bibinfo{title}{A method of computation for structural dynamics},
\newblock \bibinfo{organization}{American Society of Civil Engineers},
  \bibinfo{year}{1959}.
%Type = Book
\bibitem[{Wriggers(2008)}]{wriggers2008nonlinear}
\bibinfo{author}{P.~Wriggers}, \bibinfo{title}{Nonlinear finite element
  methods}, \bibinfo{publisher}{Springer Science \& Business Media},
  \bibinfo{year}{2008}.
%Type = Book
\bibitem[{Zienkiewicz et~al.(2000)Zienkiewicz, Taylor, Taylor, and
  Taylor}]{zienkiewicz2000finite}
\bibinfo{author}{O.~C. Zienkiewicz}, \bibinfo{author}{R.~L. Taylor},
  \bibinfo{author}{R.~L. Taylor}, \bibinfo{author}{R.~Taylor},
  \bibinfo{title}{The finite element method: solid mechanics},
  volume~\bibinfo{volume}{2}, \bibinfo{publisher}{Butterworth-heinemann},
  \bibinfo{year}{2000}.
%Type = Article
\bibitem[{Toriyama et~al.(2002)Toriyama, Tanimoto, and
  Sugiyama}]{toriyama2002single}
\bibinfo{author}{T.~Toriyama}, \bibinfo{author}{Y.~Tanimoto},
  \bibinfo{author}{S.~Sugiyama},
\newblock \bibinfo{title}{Single crystal silicon nano-wire piezoresistors for
  mechanical sensors},
\newblock \bibinfo{journal}{Journal of microelectromechanical systems}
  \bibinfo{volume}{11} (\bibinfo{year}{2002}) \bibinfo{pages}{605--611}.
%Type = Article
\bibitem[{Li et~al.(2016)Li, Gao, Chen, Yang, Li, and
  He}]{li2016nanostructured}
\bibinfo{author}{Y.~Li}, \bibinfo{author}{P.~Gao}, \bibinfo{author}{Q.~Chen},
  \bibinfo{author}{J.~Yang}, \bibinfo{author}{J.~Li}, \bibinfo{author}{D.~He},
\newblock \bibinfo{title}{Nanostructured semiconductor solar absorbers with
  near 100\% absorption and related light management picture},
\newblock \bibinfo{journal}{Journal of physics D: Applied physics}
  \bibinfo{volume}{49} (\bibinfo{year}{2016}) \bibinfo{pages}{215104}.
%Type = Article
\bibitem[{Pharr et~al.(2013)Pharr, Suo, and Vlassak}]{pharr2013measurements}
\bibinfo{author}{M.~Pharr}, \bibinfo{author}{Z.~Suo}, \bibinfo{author}{J.~J.
  Vlassak},
\newblock \bibinfo{title}{Measurements of the fracture energy of lithiated
  silicon electrodes of li-ion batteries},
\newblock \bibinfo{journal}{Nano letters} \bibinfo{volume}{13}
  (\bibinfo{year}{2013}) \bibinfo{pages}{5570--5577}.
%Type = Article
\bibitem[{Berla et~al.(2015)Berla, Lee, Cui, and Nix}]{berla2015mechanical}
\bibinfo{author}{L.~A. Berla}, \bibinfo{author}{S.~W. Lee},
  \bibinfo{author}{Y.~Cui}, \bibinfo{author}{W.~D. Nix},
\newblock \bibinfo{title}{Mechanical behavior of electrochemically lithiated
  silicon},
\newblock \bibinfo{journal}{Journal of Power Sources} \bibinfo{volume}{273}
  (\bibinfo{year}{2015}) \bibinfo{pages}{41--51}.
%Type = Article
\bibitem[{Shenoy et~al.(2010)Shenoy, Johari, and Qi}]{shenoy2010elastic}
\bibinfo{author}{V.~B. Shenoy}, \bibinfo{author}{P.~Johari},
  \bibinfo{author}{Y.~Qi},
\newblock \bibinfo{title}{Elastic softening of amorphous and crystalline li--si
  phases with increasing li concentration: a first-principles study},
\newblock \bibinfo{journal}{Journal of Power Sources} \bibinfo{volume}{195}
  (\bibinfo{year}{2010}) \bibinfo{pages}{6825--6830}.
%Type = Article
\bibitem[{Ratchford et~al.(2011)Ratchford, Schuster, Crawford, Lundgren, Allen,
  and Wolfenstine}]{ratchford2011young}
\bibinfo{author}{J.~Ratchford}, \bibinfo{author}{B.~Schuster},
  \bibinfo{author}{B.~Crawford}, \bibinfo{author}{C.~Lundgren},
  \bibinfo{author}{J.~Allen}, \bibinfo{author}{J.~Wolfenstine},
\newblock \bibinfo{title}{Young's modulus of polycrystalline li22si5},
\newblock \bibinfo{journal}{Journal of Power Sources} \bibinfo{volume}{196}
  (\bibinfo{year}{2011}) \bibinfo{pages}{7747--7749}.
%Type = Article
\bibitem[{Eidel and Ahmadi(2020)}]{eidelainsights}
\bibinfo{author}{B.~Eidel}, \bibinfo{author}{M.~Ahmadi},
\newblock \bibinfo{title}{Insights into the interplay of diffusion and stress
  in the intercalation dynamics of a li-ion battery electrode by phase-field
  finite element modeling}  (\bibinfo{year}{2020}).
%Type = Article
\bibitem[{Lee et~al.(2015)Lee, Lee, Ryu, Nix, Gao, and Cui}]{lee2015kinetics}
\bibinfo{author}{S.~W. Lee}, \bibinfo{author}{H.-W. Lee},
  \bibinfo{author}{I.~Ryu}, \bibinfo{author}{W.~D. Nix},
  \bibinfo{author}{H.~Gao}, \bibinfo{author}{Y.~Cui},
\newblock \bibinfo{title}{Kinetics and fracture resistance of lithiated silicon
  nanostructure pairs controlled by their mechanical interaction},
\newblock \bibinfo{journal}{Nature communications} \bibinfo{volume}{6}
  (\bibinfo{year}{2015}) \bibinfo{pages}{7533}.
%Type = Article
\bibitem[{Liu et~al.(2011)Liu, Huang, Picraux, Li, Zhu, and
  Huang}]{liu2011reversible}
\bibinfo{author}{X.~H. Liu}, \bibinfo{author}{S.~Huang}, \bibinfo{author}{S.~T.
  Picraux}, \bibinfo{author}{J.~Li}, \bibinfo{author}{T.~Zhu},
  \bibinfo{author}{J.~Y. Huang},
\newblock \bibinfo{title}{Reversible nanopore formation in ge nanowires during
  lithiation--delithiation cycling: An in situ transmission electron microscopy
  study},
\newblock \bibinfo{journal}{Nano letters} \bibinfo{volume}{11}
  (\bibinfo{year}{2011}) \bibinfo{pages}{3991--3997}.

\end{thebibliography}

\vfill
\newpage

\appendix

\section{The largest principal value of the effective stresses}\label{app.sig1b}
The largest principal value of the effective stresses can be obtained by
\begin{equation}
\bar{\sigma}_1 = \textrm{max} \left\{ \bm{T}_1^T \bm{\sigma}, \bm{T}_2^T \bm{\sigma} \right\} \, ,
\end{equation}
where $\bm{T}_1$ and $\bm{T}_2$ are the rotational transformation vectors which are defined as
\begin{equation}
\bm{T}_1= \left[
          \begin{array}{c}
            \cos^2 \theta \\
            \sin^2 \theta \\
            2 \sin \theta \ \cos \theta \\
          \end{array}
        \right], \qquad  \bm{T}_2= \left[
          \begin{array}{c}
            \sin^2 \theta \\
            \cos^2 \theta \\
            -2 \sin \theta \ \cos \theta \\
          \end{array}
        \right] \, ,
\end{equation}
and the rotation angle $\theta$ is obtained as follows
\begin{equation}
\tan (2 \theta) = \frac{2 \sigma_{xy}}{\sigma_{xx}-\sigma_{yy}} \, .
\end{equation}

\section{Tension part of the elastic energy density function derivatives}\label{app.tpeedfd}
Adopting the chain rule of differentiation, the tension part of the elastic energy density function derivatives with respect to $\bm u$ and $c$ respectively are derived as
\begin{equation}
\frac{\partial \left(g'(\phi) \ \xi_u^+ \right)}{\partial \bm{u}}=\frac{\partial \left(g'(\phi) \ \xi_u^+ \right)}{\partial \bm{\sigma}} \ \frac{\partial \bm{\sigma}}{\partial \bm{u}}=g'(\phi) \ \frac{\partial \left( \xi_u^+ \right)}{\partial \bm{\sigma}} \ \bm{D_1} \ \bm{B_u}^J \, ,
\end{equation}
\begin{equation}
\frac{\partial \left(g'(\phi) \ \xi_u^+ \right)}{\partial c}=\frac{\partial \left(g'(\phi) \ \xi_u^+ \right)}{\partial \bm{\sigma}} \ \frac{\partial \bm{\sigma}}{\partial c}=g'(\phi) \ \frac{\partial \left( \xi_u^+ \right)}{\partial \bm{\sigma}} \ \bm{D_2} \ N^J \, .
\end{equation}
From \ref{app.sig1b},
\begin{equation}
\xi_u^+ = \frac{1}{2E} \ \langle \textrm{max} \left\{ \bm{T}_1^T \bm{\sigma}, \bm{T}_2^T \bm{\sigma} \right\} \rangle^2 \, .
\end{equation}
Then it obtains as
\begin{equation}
\frac{\partial \left( \xi_u^+ \right)}{\partial \bm{\sigma}} = \frac{1}{E} \ \langle \textrm{max} \left\{ \bm{T}_1^T \bm{\sigma}, \bm{T}_2^T \bm{\sigma} \right\} \rangle \ \langle \frac{\textrm{max} \left\{ \bm{T}_1^T \bm{\sigma}, \bm{T}_2^T \bm{\sigma} \right\}}{\bm{\sigma}} \rangle \, .
\end{equation}

\section{Isotropic model implementation}\label{app.isomi}
If one continues \ref{eq.fipropagation} with $\xi_u$, instead of $\xi_u^+$ in hybrid formulation, one can obtain the isotropic model. Therefore, the FE implantation for isotropic model is the same, except that $R^{I}_\phi$ in \ref{eq.res} is changed into
\begin{equation}
R^{I}_\phi = \int\limits_V G_{\textrm{cr}} \ l_0 \ (\bm{B}_\phi^I)^T \ \nabla \phi \ \textrm{d}V - \int\limits_V N^I \ \left( \frac{G_{\textrm{cr}}}{l_0} \ (1-\phi) - g'(\phi) \ \xi_u \right) \ \textrm{d}V \, ,
\end{equation}
and, $\bm{K}^{IJ}_{\phi \bm{u}}$, $K^{IJ}_{\phi c}$ and $K^{IJ}_{\phi \phi}$ in \ref{eq.stif} are changed into
\begin{subequations}
\begin{align}
\bm{K}^{IJ}_{\phi \bm{u}} = \frac{\partial P^{I}_\phi}{\partial \bm{u}^J} &= \int\limits_V N^I \ g'(\phi) \ \left( \frac{1}{2} \ \bm{\varepsilon}^T \ \bm{D_1} + \frac{1}{2} \ \bm{\sigma}^T \right) \ \bm{B_u}^J \ \textrm{d}V \, , \\
K^{IJ}_{\phi c} = \frac{\partial P^{I}_\phi}{\partial c^J} &= \int\limits_V N^I \ g'(\phi) \ \left( \frac{1}{2} \ \bm{\varepsilon}^T \ \bm{D_2} \right) \ N^J \ \textrm{d}V \, , \\
K^{IJ}_{\phi \phi} = \frac{\partial P^{I}_\phi}{\partial \phi^J} &= \int\limits_V G_{\textrm{cr}} \ l_0 \ (\bm{B}_\phi^I)^T \ \bm{B}_\phi^J \ \textrm{d}V + \int\limits_V N^I \ \left( \frac{G_{\textrm{cr}}}{l_0} + g''(\phi) \ \xi_u \right) \ N^J \ \textrm{d}V \, .
\end{align}
\end{subequations}

\end{document}